\documentclass[iop]{emulateapj-rtx4} 
\shortauthors{Sekanina}
\shorttitle{Post-Perihelion Tails of Bright Sungrazers}
\slugcomment{Version \today }
\newcommand{\Rsun}{$R_{\mbox{\scriptsize \boldmath $\odot$}}\!$}

\newcommand{\lapeq}{$\;$\raisebox{0.3ex}{$<$}\hspace{-0.28cm}\raisebox{-0.75ex}{$\sim$}$\;$}

\begin{document}
\title{SPECTACULAR POST-PERIHELION TAILS OF BRIGHT KREUTZ SUNGRAZERS\\[-1.62cm]}
\author{Zdenek Sekanina}
\affil{La Canada Flintridge, California, U.S.A; {\it ZdenSek@gmail.com}
  4800 Oak Grove Drive, Pasadena, CA 91109, U.S.A.}

\begin{abstract} 
A vast majority of bright comets between the late 2nd century and the
early 18th century, moving in potentially Kreutz orbits according to
Hasegawa \& Nakano (2001), was first sighted between~2~and 16~days
after perihelion, thanks to the spectacular tails that they were then
displaying.  In~this paper I examine the basic properties
of the post-perihelion tails of the three brightest Kreutz sungrazers
of the 19th and 20th centuries --- the Great March Comet of 1843
(C/1843~D1), the Great September Comet of 1882 (C/1882~R1), and
Ikeya-Seki (C/1965~S1).  As the pre-perihelion tail of a sungrazer
sublimates completely at perihelion, the development of its post-perihelion
tail starts from scratch.  In the early days after perihelion, the
tail length grows rapidly on account of the plasma component.  At
some point the dust tail takes over, reaching a peak length weeks later.
As the geocentric distance continues to increase and the surface
brightness to decline, the tail's shortening eventually sets in.~The
dust tails of Ikeya-Seki and the 1843 sungrazer contained grains
subjected to solar radiation pressure accelerations not exceeding
0.6--0.7 the solar gravitational acceleration, the dust tail of the
1882~sungrazer was more complex.  For weeks this comet appeared
like a comet in a comet, a result of disintegration of a distant
companion near perihelion.  Evening Kreutz sungrazers are found to
have longer tails than morning ones because of geometry.  Other issues
are discussed and extensive sets of tail data are provided.
\end{abstract}
\keywords{comets general: Kreutz sungrazers; comets individual: C/1843 D1,
 C/1882 R1, C/1965 S1, C/2011 W3; methods: data analysis\\[-0.01cm]}
%

%
\section{Introduction}  
Inspection of Hasegawa \& Nakano's (2001) collection of historical appearances
of potential Kreutz sungrazing comets suggests that nearly 90~percent of these
objects recorded between the end of the 2nd century and the beginning of the
18th century were first sighted 2--16~days after perihelion, the average being
8.5~days.  Because this happens to be the time when a bright sungrazer displays
a spectacular tail that is much more conspicuous than the head, the potential
Kreutz sungrazers undoubtedly were in the given period of time detected thanks
to their post-perihelion tails.

The objective of this paper is to learn about these post-perihelion tails
by examining the tails of the brightest Kreutz sungrazers of the past two
centuries and presuming that their properties are representative of the
historical sungrazers.  The primary tasks are the discrimination between
the dust and plasma tails and the determination of a peak solar radiation
pressure acceleration that grains in the dust tails are subjected to, a
measure that governs the tail length.

\section{Post-Perihelion Tails of Sungrazers\\Ikeya-Seki and Lovejoy}
%
Very compelling evidence on the nature of the~post-perihelion tails of the
Kreutz sungrazers results from careful inspection of the observations of
comets Ikeya-Seki (C/1965~S1) and Lovejoy (C/2011~W3), the two best studied
sungrazers over the past hundred years.  However, the tail investigations
of the two objects work with greatly different data sets because these
comets were not at all alike.  Even though the nucleus of Ikeya-Seki
split near perihelion into {\it two sizable\/} fragments, the comet
survived (e.g., Marsden 1967).   On the other hand, the nucleus of
Lovejoy disintegrated about 40~hours after peri\-helion (Sekanina \&
Chodas 2012).  As a result, unlike~in the case of Ikeya-Seki (and other
surviving sungrazers), comet Lovejoy's post-perihelion (or, more precisely,
post-collapse) tail was pure dust, as no source of gas was any longer
available to replenish its ion tail.

In their investigation of Lovejoy, Sekanina \& Chodas (2012) detected
considerable sublimation of dust in close proximity of perihelion, at
heliocentric distances smaller than 1.8~solar radii, and suggested that
the dust particles in the tail were magnesium-rich olivine-based silicates,
subjected to radiation pressure accelerations of up to 0.6 the Sun's
gravitational acceleration.  This limit was also consistent with most of
the 54~post-perihelion tail length estimates collected from the period
of 2011 December 21 to 2012 March 16, with the inferred ejection times
confined to between $\sim$7~hours after perihelion and the time of
disintegration, some 30+~hours later.  Thompson (2015) measured a high
degree of polarization in Lovejoy, increasing with distance from the
nucleus and reaching as much as 58~percent or more in distant parts
of the tail; he independently invoked submicron-sized magnesium-rich
silicate grains to explain his findings.

By contrast, numerous photographic and spectral observations of the
post-perihelion tail of comet Ikeya-Seki indicated that it was a
mixed plasma-sodium-dust~feature, with a dominant optical
contribution~from~micro\-scopic dust and with plasma instabilities
responsible for superposed multiple helical structures (e.g., Krishan
\& Sivaraman 1982).

\begin{figure} 
\vspace{0.105cm}
\hspace{-0.412cm}
\centerline{
\scalebox{5.55}{
\includegraphics{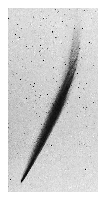}}}
\vspace{-0.05cm}
\caption{A 45-sec exposure of comet Ikeya-Seki by D.\,Milon and
 S.\,M.\,Larson, Lunar and Planetary Laboratory, with a 3.5-cm f/2.8
 camera and Tri-X panchromatic emulsion on 1965 October 30.51 UT.
 The tail was 21$^\circ$ long. (From Larson 1966.){\vspace{0.55cm}}}
\end{figure}

Figure 1 is a nice example showing the dominance of dust in the tail
of Ikeya-Seki, while the plasma features are enhanced in the tracings
in Figure~2 (Larson 1966).  The properties of the dust in the tail of
Ikeya-Seki were examined among others by Matyagin et al.\ (1968),
Weinberg \& Beeson (1976a, 1976b), Krishna Swamy (1978), Saito et al.\
(1981), and Gustafson (1985).  Using a variety of methods (polarization,
solar radiation pressure effects, microwave analogs), these authors
consistently concluded that the brightness of the post-perihelion tail
was mostly due to scattering of sunlight by microscopic silicate grains.
Matyagin et al.\ (1968) detected polarization in the dust tail in excess
of 70~percent, in fair agreement with Thompson's (2015) measurements in
the tail of comet Lovejoy nearly 50~years later.

\begin{figure} 
\vspace{0.19cm}
\hspace{-0.15cm}
\centerline{
\scalebox{0.645}{
\includegraphics{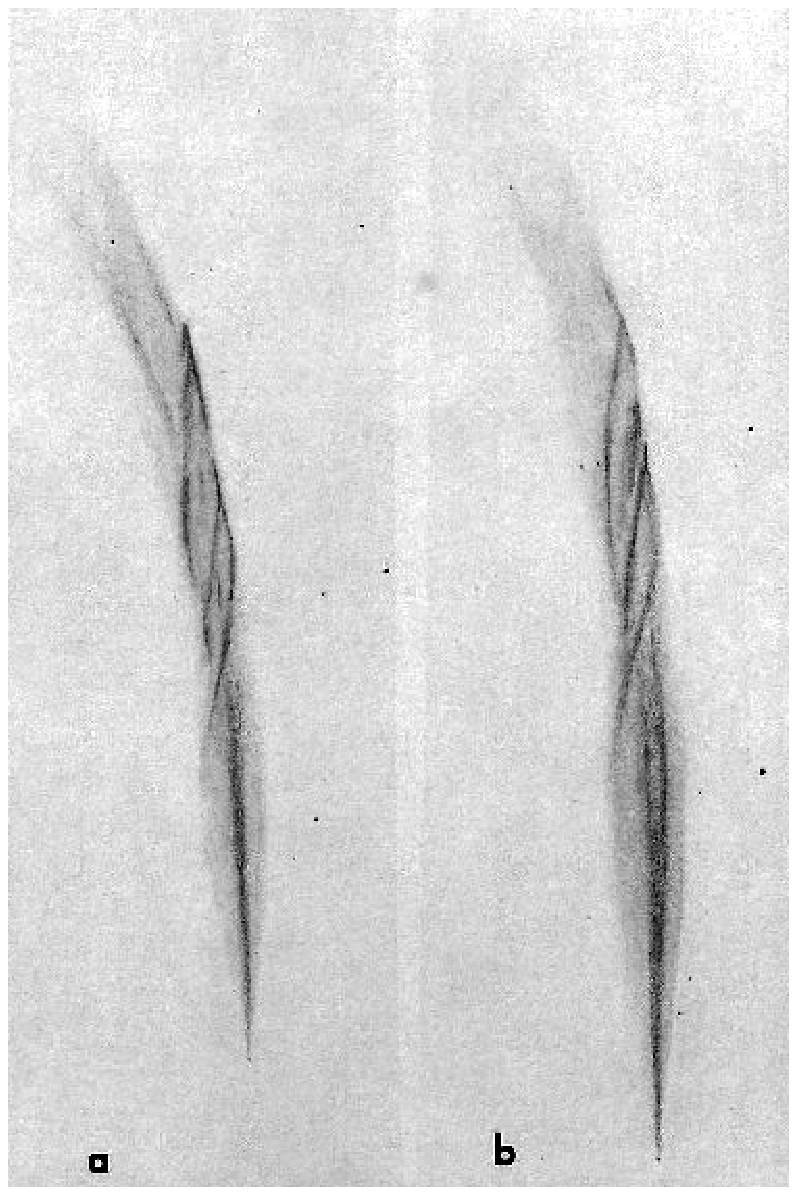}}}
\vspace{0.05cm}
\caption{Tracings of the helical structure in the tail of comet Ikeya-Seki,
 made by S.\,M.\,Larson from plates taken with the 18-cm f/7 Bailey
 astrograph of Steward Observatory; (a) October 27.52 UT, tail 15$^\circ$
 long; (b) October 28.52 UT, tail 17$^\circ$ long. North is to the left,
 west is up. (From Larson 1966.){\vspace{0.7cm}}}
\end{figure}

Milon (1969) published a comprehensive summary of the post-perihelion
tail observations of comet Ikeya-Seki made by members of the Association
of Lunar and Planetary Observers.  The list contained 80~estimates of the
tail length obtained between 1965 October~25 and December~4, some of
them visual, other measured on photographs.  Another dataset by Milon
(1969) provided information on the projected orientation of the tail's
spine between October~25 and November~6.

Besides Larson's (1966) and Milon's (1969) papers, independent data on
the post-perihelion tail of comet Ikeya-Seki were published by Antal
(1965), by Tammann (1966), and especially by Bennett \& Venter (1966),
who included the coordinates of the tail's tip.  A large fraction of
the visual observations of the length and orientation of the comet's
post-perihelion tail is available from six issues of the {\it
International Comet Quarterly\/} (Green 1982, 1985, 1987, 1991, 1993,
2001).  The grand total of observations covers a period of 85~days and
it is presented in Appendix~A:\ the tail lengths in Table~A--1, the
position angles in Table~A--2.

\subsection{Dust Tails of SOHO and STEREO\\Dwarf Kreutz Sungrazers}
I now briefly digress from the main subject of the paper to say a few
words in support of the results presented in the previous section.

Since the dwarf Kreutz sungrazers perish before reaching perihelion,
they provide no data directly relevant to the post-perihelion tails
of the bright members of the system.  However, given that all Kreutz
sungrazers are fragments of a single body, it is of interest to inspect
whether --- or to what extent --- are their basic tail properties
dependent on the fragment's size.  Two investigations suggested
that the dust in brighter, tail-displaying, dwarf sungrazers observed
with the coronagraphs on board the SOHO and STEREO space observatories
was subjected to solar radiation pressure accelerations not exceeding
$\sim$0.6 the Sun's gravitational acceleration, just as the dust in
comet Lovejoy.  This result was derived by Sekanina (2000) from
analysis of the tails of 11 SOHO sungrazers in the years 1996--1998
and independently by Thompson (2009), who triangulated the tail of a
bright dwarf comet C/2007~L3, using data from the coronagraphs on
board the two STEREO spacecraft.

\subsection{Dust in Post-Perihelion Tail of Comet Ikeya-Seki}
Even though dust was not the only constituent of the spectacular, early
post-perihelion tail of comet Ikeya-Seki, it was the primary component and
progressively the more so the farther was the comet from the Sun.  Given
that dust in the tails of comet Lovejoy and dwarf Kreutz sungrazers is
consistently subjected to the same peak radiation pressure acceleration,
one should test whether the massive amount of data on Ikeya-Seki's
post-perihelion tail length, collected in Appendix~A, is also in line
with this result.  For an assumed radiation pressure acceleration of 0.6
the solar gravitational acceleration, the {\it predicted\/} length and
orientation of this comet's tail are at several post-perihelion times
plotted in Figure~3 as a function of the ejection time.  To develop
a simple model for variations with time in the length of this sungrazer's
tail, it is first necessary to examine the relationship between the
dust tail orientation (i.e., position angle) and the ejection time of
the dust at the tail's end point.

\begin{figure}[t] 
\vspace{0.15cm}
\hspace{-0.2cm}
\centerline{
\scalebox{0.64}{
\includegraphics{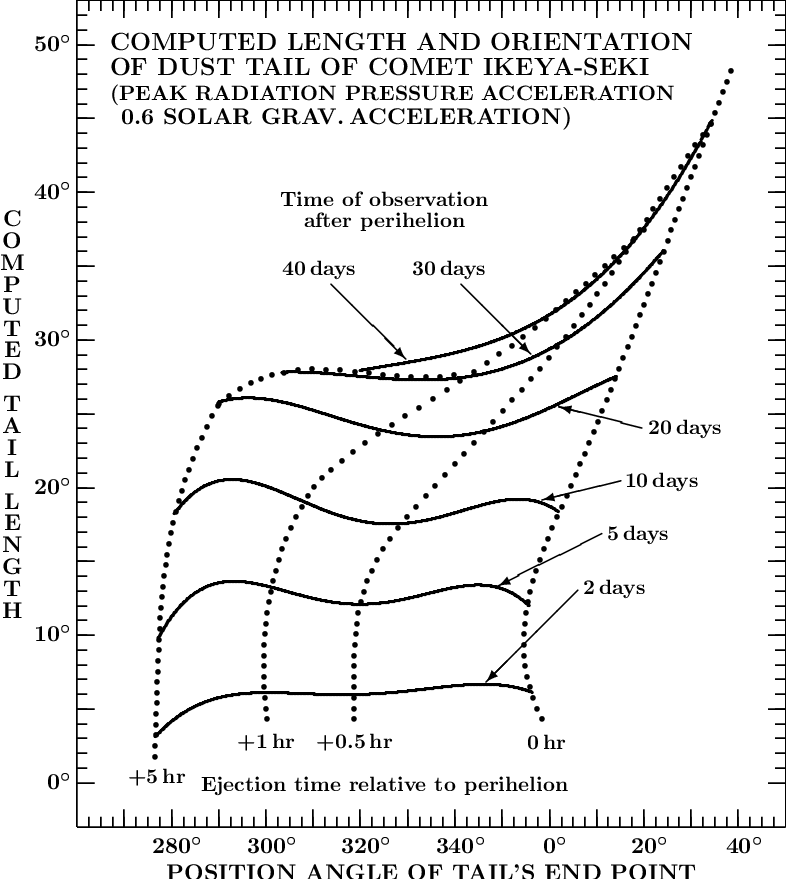}}}
\vspace{-0.05cm}
\caption{Dust tail length, computed for a peak radiation pressure acceleration
 of 0.6 the solar gravitational acceleration, against the position angle of
 the end point.  The solid curves apply to six observation times, the dotted
 curves to four ejection times.{\vspace{0.6cm}}}
\end{figure}

A list of the observed orientations of the tail of comet Ikeya-Seki is
presented in Table~A--2 of Appendix~A and the data are plotted in Figure~4.
Some of the observations made between 5 and 20 days after perihelion ---
especially the photographic ones --- show that the axis of the tail
essentially coincided with the direction of the prolonged radius vector,
an effect suggestive of a plasma tail.  On the other hand, at the times
of more than 20~days after perihelion, most of the plotted points are at
the position angles greater than the antisolar direction, implying the
presence of a dust tail.

\begin{figure}[b] 
\vspace{0.9cm}
\hspace{-0.2cm}
\centerline{
\scalebox{0.56}{
\includegraphics{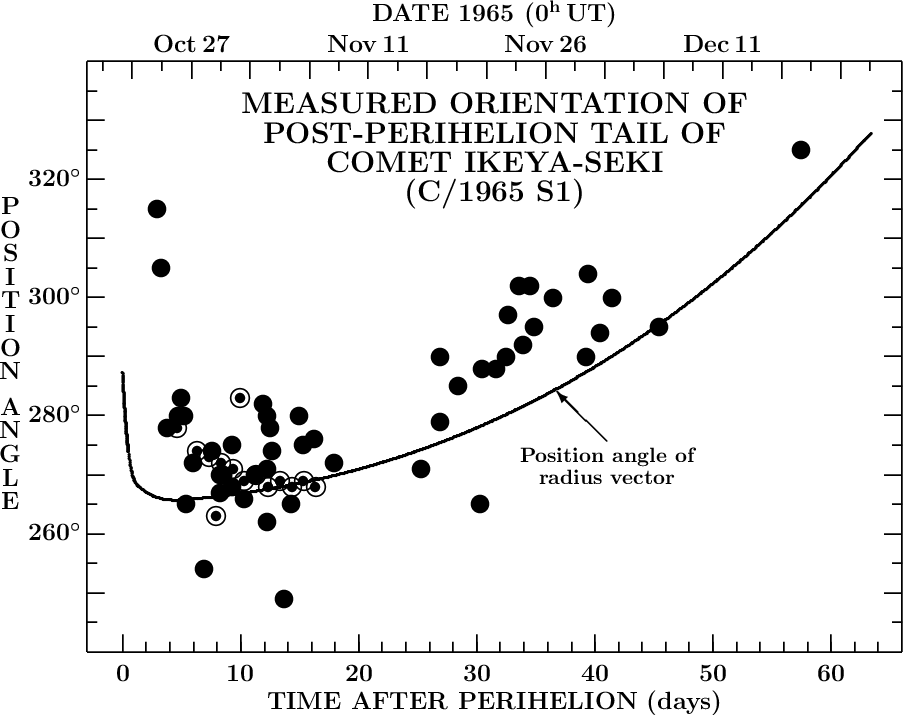}}}
\vspace{-0.05cm}
\caption{Reported position angles of the post-perihelion tail of comet
 Ikeya-Seki.  The solid circles are visual observations, the circled
 dots photographic observations.  The temporal variations in the position
 angle of the prolonged radius vector are depicted by the curve.  The
 tail orientation nearly coincided with the antisolar direction in the
 period of 5--20~days after perihelion.  Systematic deviations are
 apparent at more than 20~days after perihelion.{\vspace{-0.09cm}}}
\end{figure}
\begin{figure*}[t] 
\vspace{0.15cm}
\hspace{-0.2cm}
\centerline{
\scalebox{0.7}{
\includegraphics{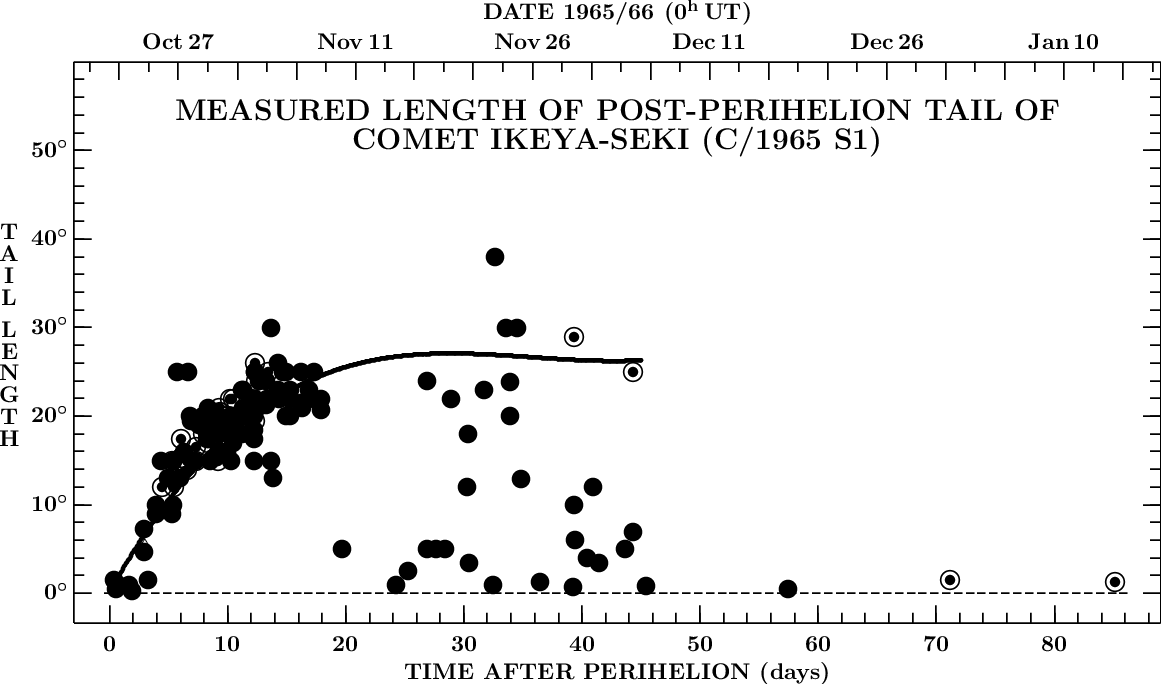}}}
\vspace{-0.05cm}
\caption{Reported lengths of the post-perihelion tail of comet Ikeya-Seki.
The solid circles are visual observations, the circled dots photographic
observations.  The predicted dependence of the dust tail length on time,
described in the text, is depicted by the thick curve.{\vspace{0.85cm}}}
\end{figure*}

Comparison of Figures 3 and 4 shows that with a possible exception of the
earliest post-perihelion observations, the dust ejecta determining the
tail length were those leaving the nucleus 5~hours after perihelion or
later.  The dust tail is predicted to have extended over approximately
10$^\circ$ on October~26, $\sim$18$^\circ$ on October~31, $\sim$23$^\circ$
on November~5, $\sim$25$^\circ$ on November~10, over not more than
27$^\circ$ on November~15, and over less than 27$^\circ$ on November~20
and beyond.

The reported post-perihelion tail lengths of comet Ikeya-Seki, summarized
in Table~A--1 of Appendix~A,~are compared with the predicted length of the
dust tail in Figure~5.  At observation times \mbox{$t_\pi < t_{\rm obs}
\leq t_\pi\!+ 45$ days} ($t_\pi$ being the perihelion time) the predicted
length $\ell$ was approximated by a polynomial
\begin{equation}
\ell = 241\!\!\:^\circ \!\cdot\tau \left(1 - 2.9 \,\tau + 2.7 \,\tau^2 \right),
  \;\;\;\; 0 \leq \tau \leq 0.45,
\end{equation}
where \mbox{$\tau = (t_{\rm obs} \!-\! t_\pi)/100$} and \mbox{$t_{\rm obs}
\!-\! t_\pi$} is in days.

The observed and predicted tail lengths are in fair agreement up to about
20~days after perihelion (i.e., November~10).  I conclude that the fundamental
assumption of a peak radiation pressure acceleration of 0.6 the solar
gravitational acceleration once again appears to fit.  As the surface
brightness of the tail was decreasing with time, nearly all reported tail
lengths farther from perihelion determined visually diminished very rapidly,
while the predicted lengths compared favorably with the lengths determined
photographically until more than 40~days after perihelion (i.e., early
December 1965), as seen from Figure~5.

\begin{figure*} 
\vspace{-0.3cm}
\hspace{0cm}
\centerline{
\scalebox{0.65}{
\includegraphics{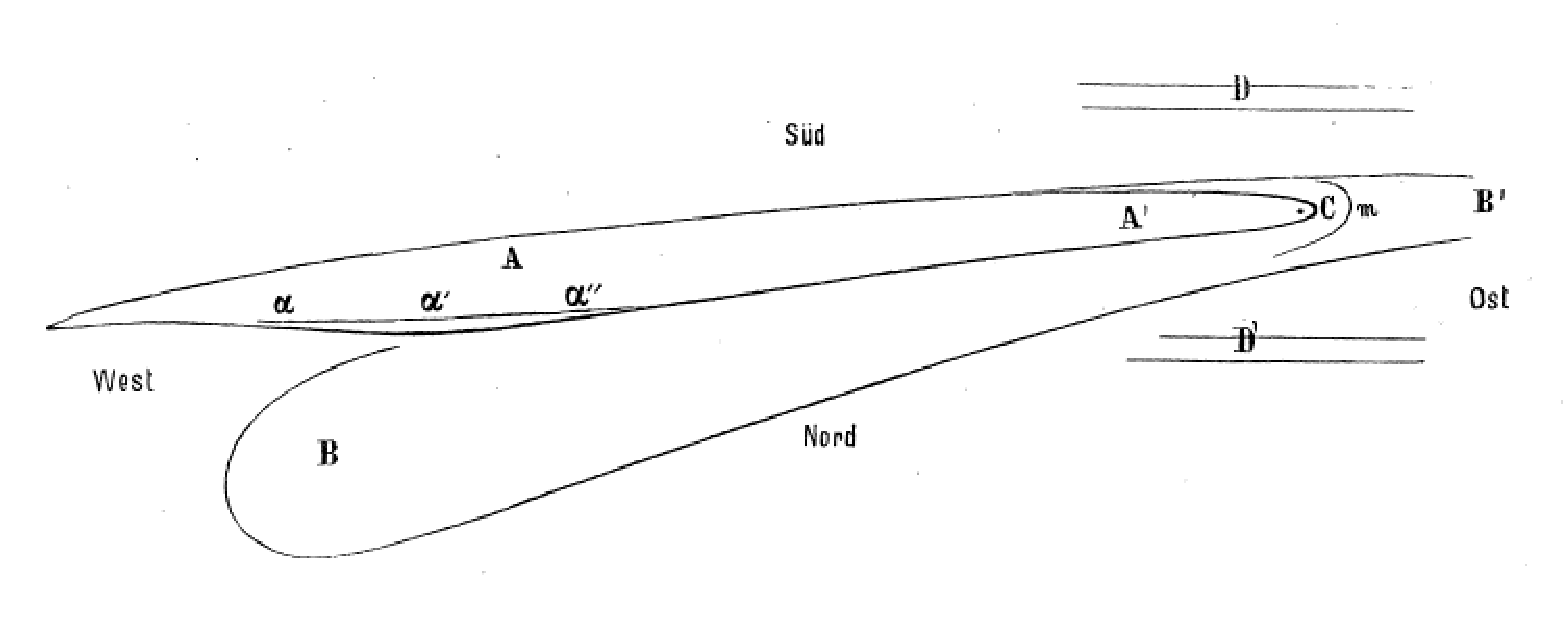}}}
\end{figure*}
\begin{figure*}
\vspace{-1cm}
\hspace{-0.14cm}
\centerline{
\scalebox{0.51}{
\includegraphics{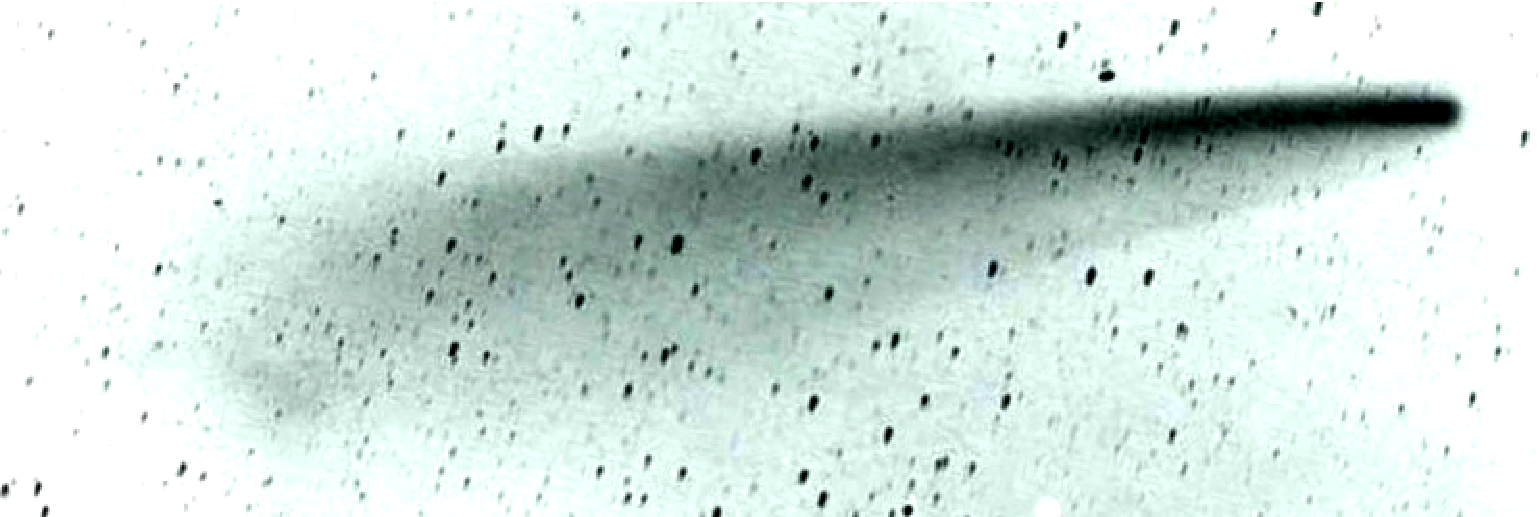}}} 
\vspace{-0.25cm}
\caption{Schmidt's (1882) schematic drawing of the Great September Comet
 after perihelion (at the top) and Gill's (1882) 100-minute exposure of
 the comet taken with a 6-cm f/4.4 camera, attached to the 15-cm Grubb
 equatorial, on 1882 November 8.06~UT.  It is noted that neither the
 drawing nor the photograph contain any featute that could be classified
 as belonging to a plasma tail; it is all dust.  The photograph has been
 scaled and rotated to fit approximately the dimensions of the drawing,
 which refers to a much earlier date and in which the north is down and
 east to the right.  The features of interest to this research are (i)~the
 ``true'' tail {\it AA}$^\prime$; (ii)~the bright spot $\alpha^\prime$,
 which is clearly resolved in the image as an isolated fuzzy cloud
 16$^\circ$ to the west-northwest from the head and which in October
 was at the middle of an elongated feature \mbox{$\alpha \: \alpha^\prime
 \alpha^{\prime\prime}$}; and (iii)~the extended region {\it
 BB\/}$^\prime$, referred to by Schmidt as a {\it nebulous tube\/}
 (Nebelrohr).{\vspace{0.8cm}}}
\end{figure*}

\section{Post-Perihelion Tail of Great September\\Comet of 1882}
%
The post-perihelion tail of the Great September Comet of 1882 (C/1882~R1)
was more complex, even though it was not reported to display the helical
structures seen early after perihelion in the tail of comet Ikeya-Seki.
To determine the type of the 1882 comet's tail, it was essential to learn
its length and orientation in the first days after the comet emerged from
the Sun's rays in late September.

As is apparent from a summary of tail length observations in Table~B--1
of Appendix~B, the first determination of this kind --- obtained visually
less than six days after perihelion --- was reported by Barnard (1884).
Although he did not nominally measure the tail's position angle, he did
remark that the tail --- very narrow, only about 1$^\circ\!$.5 wide at
most, with the boundaries sharply defined --- made at the time an angle
of 45$^\circ$ with the horizon.  This information allows one to derive
the tail's position angle of 262$^\circ$, within 1$^\circ$ of the radius
vector.  The tail was 15$^\circ$ long at the time, and this requires a
radial acceleration of about 10~times the solar gravitational acceleration,
far beyond the range of dust-grain accelerations, but typical for the
plasma tails.  At the same time, Barnard also remarked that the tail was
slightly convex to its south side, a feature that is characteristic of
a dust tail.  An obvious conclusion is that, early after perihelion, the
1882 comet too appeared to exhibit a mixed, dust-plasma tail.

Gill (1882) took six photographs of the comet and its tail with a small
camera at the Royal Observatory at Cape between 1882 October~20 and
November~15~UT; the exposure times were 30 to 140~minutes.  However,
the photographs appear to have never been examined and no results were
published for decades (Gill 1911).

Visually, the post-perihelion tail was observed rather extensively.
Five dedicated studies were undertaken by, respectively, Schmidt (1882,
1883) at Athens; Schwab (1883) at sea (on board the ship Thebes);
Barnard (1883, 1884) at Nashville, Tenn.; E.\,Frisby, A.\,N.\,Skinner,
and W.\,C.\,Winlock at the U.S. Naval Observatory (Winlock 1884); and
Leavenworth \& Jones (1914) at the Leander McCormick Observatory.
Observations were also secured by Andr\'e (1882), Ellery (1882a,
1882b), Galle (1882), Kortazzi (1882), Ledger (1882), Palisa (1882),
von Engelhardt (1882, 1883), Backhouse (1883), and Gould (1883).  The
in-depth report by Leavenworth \& Jones includes a number of the tail's
drawings as well as their extensive description.

\subsection{Schmidt's Observations of a Bright Spot $\alpha\!\:^\prime$}
Schmidt (1882) remarked that the ``true'' tail, which in the upper part
of Figure~6 occupies the general region {\it AA}$^\prime$, was up to
23$^\circ$ long and terminated in a faint, sharp tip.  He further noted
that parallel to this tail ran a ``nebulous tube'' (Nebelrohr) {\it
BB\/}$^\prime$ that appeared to emanate from a feature located several
degrees from the comet's head, {\it C\/}, sunward (to the east) of it.
This feature was also reported by other observers and extensively
described by Schwab (1883); see Section~3.3.  In addition to two
branches of light, {\it D\/} and {\it D\/}$^\prime$, and a temporary
halo, {\it m\/}, Schmidt also noted, near the northwestern end of the
true tail, a strikingly bright ``hem'' (Saum), \mbox{$\alpha \,
\alpha^\prime \alpha^{\prime\prime}$}.  From November~6 on, the
hem's bright spot $\alpha^\prime$ grew into a distinct, isolated
0$^\circ\!$.5 wide cloud, which is clearly visible in Gill's 100-minute
exposure in Figure~6.

Schmidt provided approximate equatorial coordinates of the feature
$\alpha\!\:^\prime$, for the equinox of 1850.0, on 24~occasions between
October~6 and November~17~UT, with the times given to 0.1~hr.  By
converting the coordinates to the equinox of J2000.0 and by computing
the topocentric ephemeris for the nucleus at the given times, one can
examine the motion of this feature through the tail.  The positions as
a function of time should provide constraints on both the ejection time
and the effective solar radiation pressure that the dust grains making
up the feature were subjected to.

The results of the modeled feature are presented in Table~1.  Column~2
lists the cometocentric latitude of the Earth, which provides information
on the degree of resolution with which we view the motions of features,
such as $\alpha\!\:^\prime$, in the orbital plane of the comet.  No
modeling would effectively be possible, if the angle should be very
close to zero, because our view would then be essentially edgewise.
Columns~3 and 4 are the observed polar coordinates of the feature,
derived from Schmidt's original data after they were converted to the
equinox of J2000.  The last four columns offer the results of the
solution that was deemed to provide an optimized approximation, but
not a least-squares solution.

Ignoring a potential effect of a separation velocity, the presented
fit was obtained for an ejection time, $t_{\rm ej}$, of 0.9~day after
perihelion and an effective radiation pressure acceleration,
$\beta_{\rm eff}$, of 0.7 the solar gravitational acceleration.  The
mean residuals were $\pm$0$^\circ\!$.82 in the distance from the
nucleus and $\pm$1$^\circ\!$.62 in the position angle.  However,
even at a radial distance of 17$^\circ$, slightly exceeding the peak
tabulated distance, the mean residual in the position angle would be
equivalent to a transverse distance of only $\pm$0$^\circ\!$.48; the
uncertainty in the radial distance thus clearly dominated.  As seen
from Figure~6, the diameter of the feature was about 0$^\circ\!$.5 on
November~8, so that the uncertainty exceeded it by a factor of nearly
two.  Under these circumstances, one should accept the results of
modeling with caution.

\begin{table*}[t]  
\vspace{0.15cm}
\hspace{-0.15cm}
\centerline{
\scalebox{1}{
\includegraphics{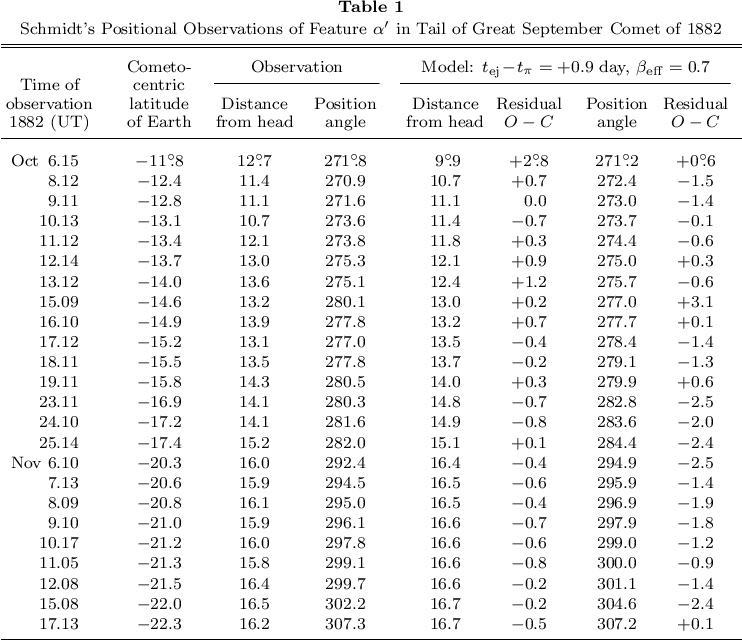}}}
\vspace{0cm}
\end{table*}
\begin{table*}[t]  
\vspace{0.1cm}
\hspace{-0.15cm}
\centerline{
\scalebox{1}{
\includegraphics{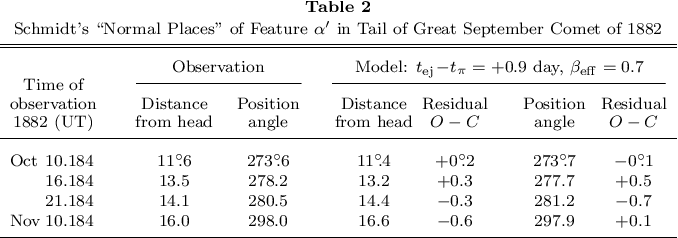}}}
\vspace{0.8cm}
\end{table*}

As Schmidt tried to investigate the feature's motion himself, he may
have noticed how incongruous his measurements were.  This may have
been the reason for his introduction of four ``normal places'' by
averaging the 24~points.  I compare these more representative
positions with the same model solution in Table~2.  The mean
residuals do indeed improve significantly, amounting to
$\pm$0$^\circ\!$.44 in the distance from the nucleus and
$\pm$0$^\circ\!$.50 in the position angle, equivalent to
$\pm$0$^\circ\!$.12 in the angular distance in the transverse
direction.

The obvious difference between the dust grains populating the most
distant parts of Ikeya-Seki's tail and the dust that made up the
feature $\alpha\!\:^\prime$ in the tail of the 1882 comet is in that
the latter was ejected from the nucleus later after perihelion than
the former.  The difference in the effective radiation pressure
acceleration appears to be minor.  In the context of these inconclusive
results it is desirable to examine the tail length and orientation
of the 1882 comet.

\subsection{Tail Length and Orientation of the 1882 Sungrazer}
It is unfortunate that information on the tail orientation of the
Great September Comet of 1882 is extremely limited; the set of
simultaneous data on the tail's orientation and length is downright
pitiful, ruling out application of the approach employed for
Ikeya-Seki.  Instead I used the pairs of the tail length and
position angle to determine the correponding pairs of the ejection
time and radiation pressure acceleration, as shown in Table~3.

\begin{table}[b] 
\vspace{0.6cm}
\hspace{-0.25cm}
\centerline{
\scalebox{1}{
\includegraphics{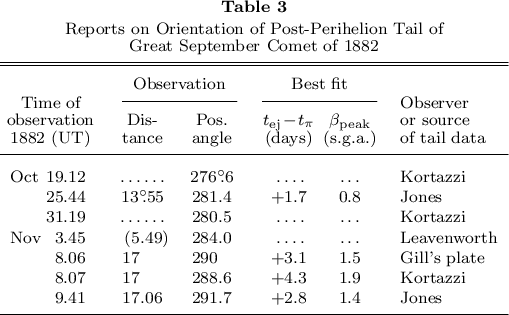}}}
\vspace{0cm}
\end{table}
\begin{figure*}[t] 
\vspace{0.15cm}
\hspace{-0.55cm}
\centerline{
\scalebox{0.7}{
\includegraphics{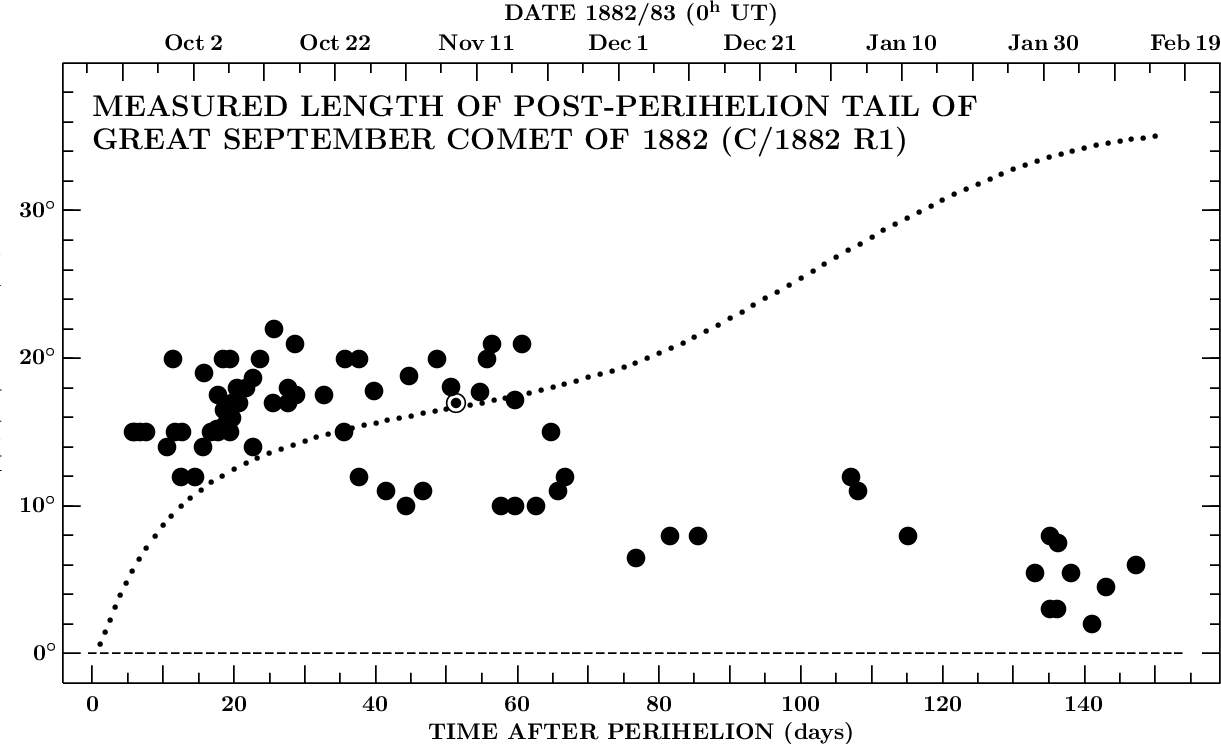}}}
\vspace{-0.05cm}
\caption{Reported lengths of the post-perihelion tail of the Great
 September Comet of 1882 from visual observations (solid circles).
 The circled dot is the single data point based on my estimate of
 the tail length from Gill's photograph in Figure~6.  The dotted
 curve shows the dust tail lengths that satisfy the ejection
 conditions employed for comet Ikeya-Seki in Figure~5.{\vspace{0.59cm}}}
\end{figure*}

The position angles listed in the table came from three sources:\ the
observations by Kortazzi (1882) at Nikolayev and by Leavenworth \& Jones
(1915) at the Leander McCormick Observatory in Virginia were supplemented
by my estimate from Gill's plate in Figure~6.  Only four of the seven
entries were suitable for analysis, as the other three did not include
the tail length or its realistic estimate.  Yet, the few numbers suffice
to demonstrate significant differences between the 1882 comet and Ikeya-Seki.
The particulates defining the length of the 1882 comet's tail were released
from the nucleus at times that lagged the perihelion passage a few days
rather than a small fraction of a day and were subjected to much higher
radiation pressure accelerations.  If correct, this result and an apparent
correlation between both parameters in Table~3, imply that unless the
reported tail lengths referred to a plasma tail --- which is highly
unlikely --- the dust in the 1882 comet included absorbing grains,
contrary to Ikeya-Seki, which contained only dielectric particles.

Similar conclusions are reached from the plot of the 1882 comet's tail length
against the observation time in Figure~7.  Although straightforward comparison
with Figure~5 indicates that the observed length of Ikeya-Seki's tail was in
fact longer than that of the 1882 sungrazer, the difference was entirely an
effect of projection.  Figure~7 demonstrates that in the first 30~days after
perihelion the tail of the 1882 comet was systematically longer than it should
have been, if under the ejection conditions equivalent to those for comet
Ikeya-Seki.  Interestingly, the tail length estimated from Gill's photograph
in Figure~4 fits the Ikeya-Seki curve just about perfectly.  The problem is
a difference of tens of degrees in the position angle.  And even though, in
general, Table~3 and Figure~7 allow either interpretation, Gill's photograph
shows no trace of a plasma tail.

In any case, an obvious question is why are the tails of the two sungrazers ---
undoubtedly fragments of a common parent --- so different?  It is conceivable
that the culprit was the extensive, multiple fragmentation of the nucleus
of the 1882 comet at perihelion.  Although Ikeya-Seki split as well, only
two persisting components were observed after perihelion.  By contrast, the
Great September Comet broke up into as many as six major active fragments and
the comet's post-perihelion light curve is known to have been significantly
less steep than Ikeya-Seki's, $r^{-3.3}$ vs $r^{-3.9}$ (Sekanina 2002).  The
comprehensive fragmentation was likely to bring to the surface the material
that otherwise would have stayed hidden in the interior, and the flatter
light curve suggests a more substantial contribution to the comet's activity
from the ejecta released farther from perihelion.

\begin{figure*} 
\vspace{-0.3cm}
\hspace{0cm}
\centerline{
\scalebox{0.62}{
\includegraphics{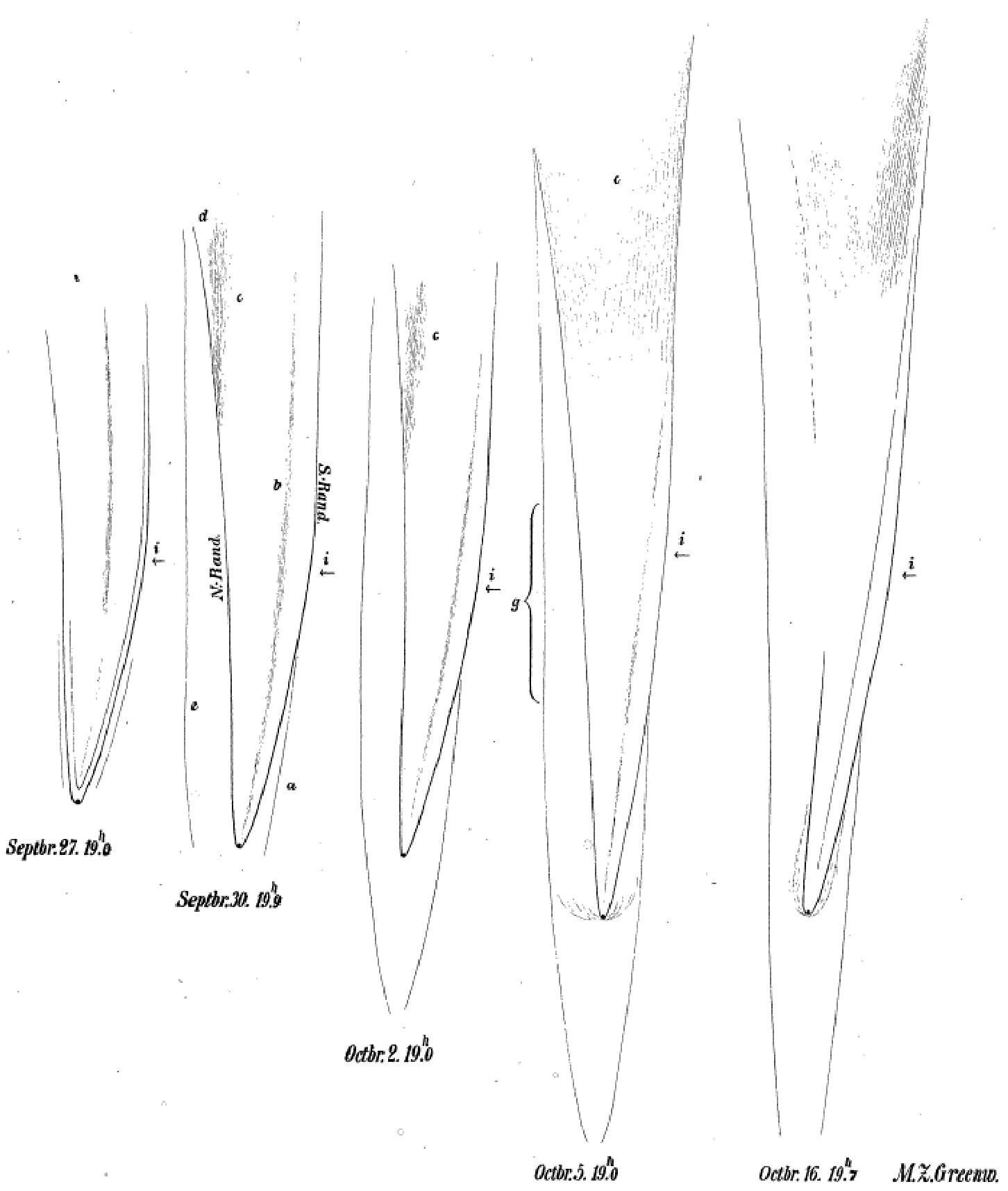}}}
\vspace{-0.4cm}
\caption{Drawings of the Great September Comet of 1882 by Schwab (1883),
 who referred to Schmidt's nebulous tube as {\it H\"{u}lle\/}, i.e.,
 {\it sheath\/} or {\it envelope\/}.  Note that the features make an
 impression of a comet in a comet.  Schwab warned that his drawings
 were not to the same scale and pointed out that on October 17 the
 sheath extended over 5$^\circ$--\,6$^\circ$ sunward of the comet's
 head in an opera glass.{\vspace{0.6cm}}}
\end{figure*}

\subsection{A Comet in a Comet}
The feature that Schmidt called the nebulous tube is particularly
intriguing.  This term is not very fitting and even though Schwab's
is better, the impression that I am getting when looking at their
drawings is that of a {\it comet inside another comet\/}.  The inner
comet is the Great September Sungrazer itself, which I will in this
section refer to as the {\it main comet\/}.  The other, with no
nuclear condensation and no sunward boundary, I will call the
{\it outer comet\/}.  I am unaware of any other cometary object
ever reported to possess such an unusual appearance.

The outer comet must have had a nuclear condensation and sunward
boundary at some point in the past.  Their loss is a key piece of
evidence for the scenario proposed below.  I begin with the
angular distance between the nucleus of the main comet and the site
of the missing nucleus of the outer comet, which is assumed to be
closely approximated by the separation of the observed sunward end
of the outer comet from the nucleus of the main comet.  On Schmidt's
drawing in Figure~6 this separation is marked as a distance
{\it B\/$^\prime\!$C\/}, for which both observers offered their
estimates.  The important difference was that Schmidt (1882)
provided only a general comment, saying that the {\it B\/$^\prime\!$C\/}
part of the nebulous tube, seen until November~21, was 1$^\circ$ wide and
3$^\circ$--\,5$^\circ$ long in a seeker telescope; and that it was only
1$^\circ$--\,2$^\circ$ long to the naked eye when the moonlight did not
interfere.  Schwab (1883), on the other hand, estimated the separation at
5$^\circ$--\,6$^\circ$ in an opera glass and 3$^\circ$--\,4$^\circ$ with the
naked eye on October~17.3~UT.  This is important because calculations show
that the separation {\it B\/$^\prime\!$C\/} was a strong function of time.
Following Schwab, I adopted that the separation between the nuclei equaled
5$^\circ\!$.5 on October~17.3~UT.

\begin{table*}[t] 
\vspace{0.2cm}
\hspace{-0.2cm}
\centerline{
\scalebox{1}{
\includegraphics{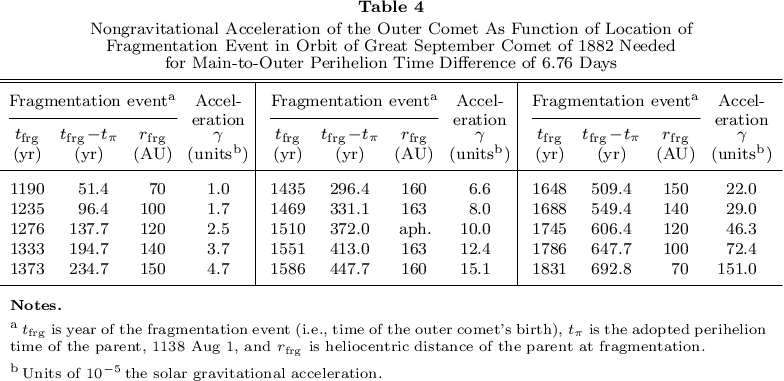}}}
\vspace{0.63cm}
\end{table*}

Next I assumed that the outer comet followed the main comet in the
latter's orbit about the Sun, and from the ephemeris I computed the
difference in the perihelion time needed to change the orbital
position by 5$^\circ\!$.5 along the projected orbit on October~17:\
the result was that the outer comet must have passed the perihelion
point 6.76~days {\it after\/} the main comet.

It was now time to propose a scenario for the outer comet, which
involved the parent of the 1882 sungrazer --- the Chinese comet of
1138 (Sekanina \& Kracht 2022) --- but was very different from the
scenario for comet Ikeya-Seki (Marsden 1967), as one should expect.
On the other hand, since the outer comet passed the 1882 perihelion
nearly a week after the main comet, the separation must have taken
place {\it very long\/} before that perihelion (see Sekanina 1982).
To survive, the fragment ought to have been fairly massive (although
much less massive than the nucleus of the main comet) and subject
to a relatively low sublimation-driven nongravitational
acceleration, of $\!\!${\lapeq}$\!\!$10 units of 10$^{-5}$\,the
solar {\vspace{-0.03cm}}gravitational acceleration (in the following:\
``units'').  A fragment of that kind --- here the outer comet ---
orbits the Sun in a gravity field slightly weaker than is the Sun's
gravitational field and, accordingly, its motion is marginally slower
than the motion of the principal fragment ---here the main comet ---
hence, the {\it lag\/}.  The rate of this slowdown in terms of an
orbital-period effect, $\Delta P$, can in the {\it absence of any
momentum exchange at breakup\/} be estimated from a simple equation:
\begin{equation}
\Delta P = {\textstyle \frac{1}{2}} \,10^{-5}\gamma P,
\end{equation}
where $\gamma$ is the nongravitational acceleration (in units) and $P$ is
the orbital period of the parent comet ($P$ and $\Delta P$ in the same
units).  Adopting for the outer comet \mbox{$\gamma${\lapeq}10 units}
and for the parent{\vspace{-0.05cm}} sungrazer \mbox{$P = 744$ yr} (1882
minus 1138), one obtains \mbox{$\Delta P${\lapeq}14 days}.  More specific
conditions are found from the dependence of the nongravitational
acceleration $\gamma$ that implies an 1882 perihelion time difference of
6.76~days as a function of the fragmentation time.  These results are
presented in Table~4.  It follows that the breakup took place most
probably on the way to aphelion, not later than in the early 16th
century.  Otherwise the implied nongravitational acceleration suggests
that the outer comet would have been too small to survive intact to the
1882 perihelion.

If the outer comet was involved in any momentum exchange at the time of
separation, the numbers in Table~4 should be replaced by another infinite
number of possibilities that satisfy the condition of the outer comet
lagging 6.76~days behind the main comet at the 1882 perihelion.  In one
such scenario, with potentially important implications, the outer comet
could have separated from the parent at aphelion in 1510 with a velocity
whose transverse component was 3~m~s$^{-1}$ in the direction opposite
the orbital motion and radial component 0.1~m~s$^{-1}$ in the antisolar
direction, and then be subjected to a nongravitational acceleration of
5.7~units.  Besides arriving at the 1882 perihelion 6.76~days after the
main comet, the outer comet would do so in an orbit whose perihelion
distance dropped from 1.67\,{\Rsun} (at a zero separation velocity) to
1.25\,{\Rsun}\,.

Another set of conditions that the outer comet is expected to satisfy
concerns its physical behavior around the 1882 perihelion.  The object
was either overlooked or too faint to detect before perihelion.  Very
shortly after perihelion, within several hours or so, the entire body
suddenly exploded or collapsed, its mass disintegrating into dust of
a wide particle size distribution.  The nucleus with its condensation
disappeared and much of the dust cloud on the sunward side completely
sublimated --- especially if the perihelion distance dropped as suggested
above --- so that in a moment the object gained the appearance consistent
with the observations.  The dust cloud on the antisunward side survived
and instantly began to form the tail.  This proposed development is no
fairy tale, because at least two sungrazers are known to have disintegrated
instantaneously near perihelion.  One was comet Lovejoy (C/2011~W3),
which suddenly collapsed about 40~hours after perihelion, completely
changing its appearance from day to next day (Sekanina \& Chodas 2012);
the other, apparently smaller object, was the Great Southern Comet of
1887 (C/1887~B1), which is believed to have instantly perished
$\sim$6~hours after perihelion (Sekanina 1984).  Sizewise the outer
comet perhaps was in between the two, as it was observed for a little
longer than two months following the apparent time of disintegration
(see below).  Lovejoy was visible for three months and the 1887
sungrazer for barely three weeks.:

Next I examine the timeline of the outer comet.  In the adopted scenario
the object's perihelion passage occurred on October~24.5~UT and it
disintegrated several hours later, as will be apparent from the following.
A reported tail observation of the outer comet on or before this date
would invalidate the hypothesis.  Schmidt (1882) observed this feature
from October~4 to November~21.  Although Schwab (1883) commented on
two envelopes on September~24 and 27 (i.e., 25.3 and 28.3~UT), his
drawings from five mornings between September~28.3~UT and October~17.3~UT
(shown in Figure~8) tell a different story.  In the first picture, from
September~28.3, the main comet is seen to display {\it three\/} envelopes,
very close to each other, around {\it its\/} nucleus, located at the tip
of the condensation; there is no trace of any feature sunward of it.  Only
the second drawing, from October~1.3~UT, shows an outer envelope, much
wider than the main envelope and with the axis shifted slightly to the left
(towards the north).  The three remaining sketches copy the layout of the
second picture, including the asymmetry.  However, the sunward extension
of the outer envelope is drawn nearly to the tip, especially on the
drawings from October~3.3 and 6.3, on which nothing stands in the way
to the reader's overwhelming perception of these features as {\it a comet
in a comet\/}.  Schwab (1883) indicated that he observed the outer comet
from October~1.3 to November~13.2~UT (on this last date from Punta Arenas,
Chile), when he reported that the object's figure was still looking like
on October~17.  The periods of time over which Schmidt and Schwab observed
the outer comet were then fairly comparable.

Examining the observed separation between~the~main comet's nuclear
condensation and the computed position of the outer comet's disintegrated
nucleus was high priority.  This distance varied from 11$^\circ$ on
September~28.3 to Schwab's averaged 5$^\circ\!$.5 on October~17.3~UT,
so that the outer comet was gradually catching up with the main comet.
Interestingly, both Schmidt's and Schwab's drawings show the tails of
the main and outer comets extending to just about equal distances from
the Sun in projection onto the sky, a trait also plainly visible in
Gill's November~8 photograph in Figure~6.  Given that the location of
the outer comet's disintegrated nucleus was degrees sunward of the main
comet's nuclear condensation, the deficit had to be made up by the outer
comet's longer projected tail.  At first sight, this looks like a difficult
condition to satisfy.

To examine this problem in detail, I use the distance of the feature
$\alpha\,^\prime$ from the nucleus of the main comet as a measure of
this comet's tail length, as $\alpha\,^\prime$ indeed was located near
the tail's very end (Figure~6).  This definition of the tail length is
convenient, because the position of $\alpha\,^\prime$ relative to the
nucleus is tightly determined by the dynamical parameters (see Tables~1
and 2).  Now, let at time $t$ the position vector of the tail's end
point relative to the nucleus be {\bf L}$_{\rm I}(t)$.  If the position
vector of the nucleus relative to the Sun at time $t$ is {\bf E}$_{\rm
I}(t)$, the position vector of the tail's end point relative to the Sun
equals \mbox{{\bf E}$_{\rm I}^\ast(t)$ = {\bf E}$_{\rm I}(t)$\,+\,{\bf
L}$_{\rm I}(t)$}.  The position vector {\bf E}$_{\rm I}(t)$ projects
onto the plane of the sky as a solar elongation of the main comet's
nucleus, $E_{\rm I}(t)$, while the position vector {\bf E}$_{\rm I}^\ast(t)$
as a solar elongation of the tail's end point, $E_{\rm I}^\ast(t)$.

\begin{table}[t] 
\vspace{0.14cm}
\hspace{-0.19cm}
\centerline{
\scalebox{0.985}{
\includegraphics{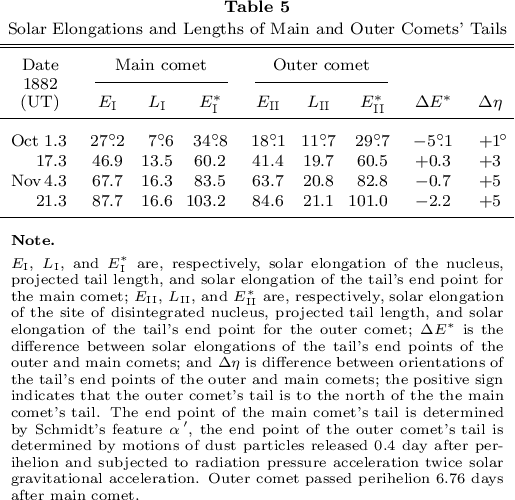}}}
\vspace{0.7cm}
\end{table}

Let, similarly, the position vector of the disintegrated nucleus of the
outer comet relative to the Sun at time $t$ be {\bf E}$_{\rm II}(t)$.
Since the peak solar radiation pressure on dust particles in the tail
of the 1882 sungrazer was near \mbox{$\beta_{\rm peak} \simeq 2$} units
of the solar gravitational acceleration (Table~3), I assume that this
too was the peak radiation pressure acceleration that dust particles
released from the outer comet were subjected to.  And since the ejection
of all dust occurred at once, the motion of the particles with $\beta_{\rm
peak}$ determined the tail length of the outer comet.  If the position
vector of the tail's end point relative to the location of the disintegrated
nucleus at time $t$ is {\bf L}$_{\rm II}(t)$, its position vector relative
to the Sun equals \mbox{{\bf E}$_{\rm II}^\ast(t)\!=\,${\bf E}$_{\rm II}(t)\:\!
+ $\,{\bf L}$_{\rm II}(t)$}.  The lengths of the position vectors
{\bf E}$_{\rm II}(t)$ and {\bf E}$_{\rm II}^\ast(t)$ project onto the
plane of the sky as solar elongations of, respectively, the outer comet's
disintegrated nucleus, $E_{\rm II}(t)$, and its tail's end point,
$E_{\rm II}^\ast(t)$.

The choice of the time of the outer comet's disintegration determines
the shift between the tails of the main and outer comets:\ the earlier
the time the farther to the north would the outer comet's tail be moved
and the greater would be the angle subtended by the two tails' axes.
In space this angle, $\Delta \epsilon$, is measured by the scalar product
of the two normalized vectors,
\begin{equation}
\Delta \epsilon(t) = \arccos \! \left[\frac{{\bf E}_{\rm I}^\ast(t)
 \,\mbox{\Large \boldmath $\cdot$}\, {\bf E}_{\rm II}^\ast(t)}{|{\bf
 E}_{\rm I}^\ast(t)| \!\cdot\! |{\bf E}_{\rm II}^\ast(t)|} \right] \!, 
\end{equation}
and in projection it shows up as a position angle difference $\Delta
\eta$; it is positive when the outer comet's tail is to the north of
the main comet's tail and vice versa.  The observations (Figures~6
and 8) consistently show that $\Delta \eta$ was positive but very
small, at most several degrees.

Experimentation with the syndyname approach suggested that to fit
these conditions on $\Delta \eta$, the disintegration of the nucleus
of the outer comet ought to have taken place about 0.4~day after its
perihelion passage, i.e., on September 24.9 UT.  A solution is
presented in Table~5, which shows that, except in early October,
the solar elongations of the two tails' end points, given by~$\Delta
E^\ast$, indeed came out to be about equal.  The tails' angular
lengths are listed in the table as $L_{\rm I}$ and $L_{\rm II}$.
Note that because of the projection effects \mbox{$E_x \!+\! L_x
\geq E_x^\ast$} (for \mbox{$x = \;$I, II}).

This exercise demonstrates that the hypothesis of {\it a comet in
a comet\/} is plausible, as it is fully supported by the
relevant computations.

As the last point I comment on an estimated width of the outer
comet's tail and the ramifications for the particle ejection
velocities.  Schmidt (1882) indicated that the tail was about
1$^\circ$ wide, but provided no details.  An assumption that
the ejection velocity alone accounts for the tail's width leads
to a crude upper limit of \mbox{1--2 km s$^{-1}$}.  More realistic
estimates are a factor of several lower, because the broad range
of radiation pressure accelerations contributes significantly to
the tail's width as well.  Accordingly, the vague data on the
velocity of ejecta are not critical and should introduce no major
obstacles for accepting the proposed hypothesis.

\section{Post-Perihelion Tail of Great March\\Comet of 1843}
%
This sungrazer is believed to have displayed~one~of~the longest and
most impressive tails that have ever been observed.  I was able to
secure a surprisingly large number of estimates of this tail's length
by inspecting~the~\mbox{major} journals of the time.  Data were
contributed by~Cooper (1843), J.\,G.\,Galle (Schumacher 1843),
Haile (1843),~Kay (1843), Knorre\,(1843), von Littrow\,(1843),
Tucker\,(1843), Brand (1844), Maclean (1844), E.\,Dunkin and
J.\,Glaisher (Airy 1845), and Caldecott (1846).

\subsection{Observations by C.\ Piazzi Smyth at Cape}

A dilligent observer of the comet and major~\mbox{contributor} to the pool of
available data was C.\ Piazzi Smyth, the first assistant to Director of the
Royal Observatory, Cape of Good Hope.  When the comet suddenly appeared in
close proximity of the Sun at the end of February and in early March, he was,
at the age of 24, the only astronomer at the Observatory (Warner 1980).
While complaining about the inadequate instrumentation, he managed to secure
extensive data on the comet, including numerous drawings of its changing
appearance during March.  He must have been working extensively on assembling
the results --- including 11~drawings made between March~3 and 31, five
naked-eye views and six telescopic --- because he sent them in a letter to
the {\it Monthly Notices of the Royal Astronomical Society\/} nearly three
years after the observations were made (Piazzi Smyth 1846).  Unfortunately,
only a short extract of the accompanying letter was published by the journal's
editors and no pictures (judging from the ADS scans).

It was Warner's (1980) publication of detailed extracts from Piazzi Smyth's
diary that furnished the most welcome data on the tail of the Great March
Comet.  Included were, in particular, the equatorial coordinates of the end
of the tail on nine days between 1843 March~3 and 22, which I combined with
a single estimate of the length and orientation by Knorre (1843) on March~17
to examine the comet's tail, using the technique applied to Schmidt's
observations of the Great September Comet of 1882.  Thanks to these data,
I did not have to employ J.\,F.\,W.\,Herschel's reports on stars passing
through the tail (e.g., Schumacher 1843; Kapoor 2021), from which the tail
orientation should at best be determined only very approximately.

\begin{figure}[t] 
\vspace{0.16cm}
\hspace{-0.15cm}
\centerline{
\scalebox{0.6}{
\includegraphics{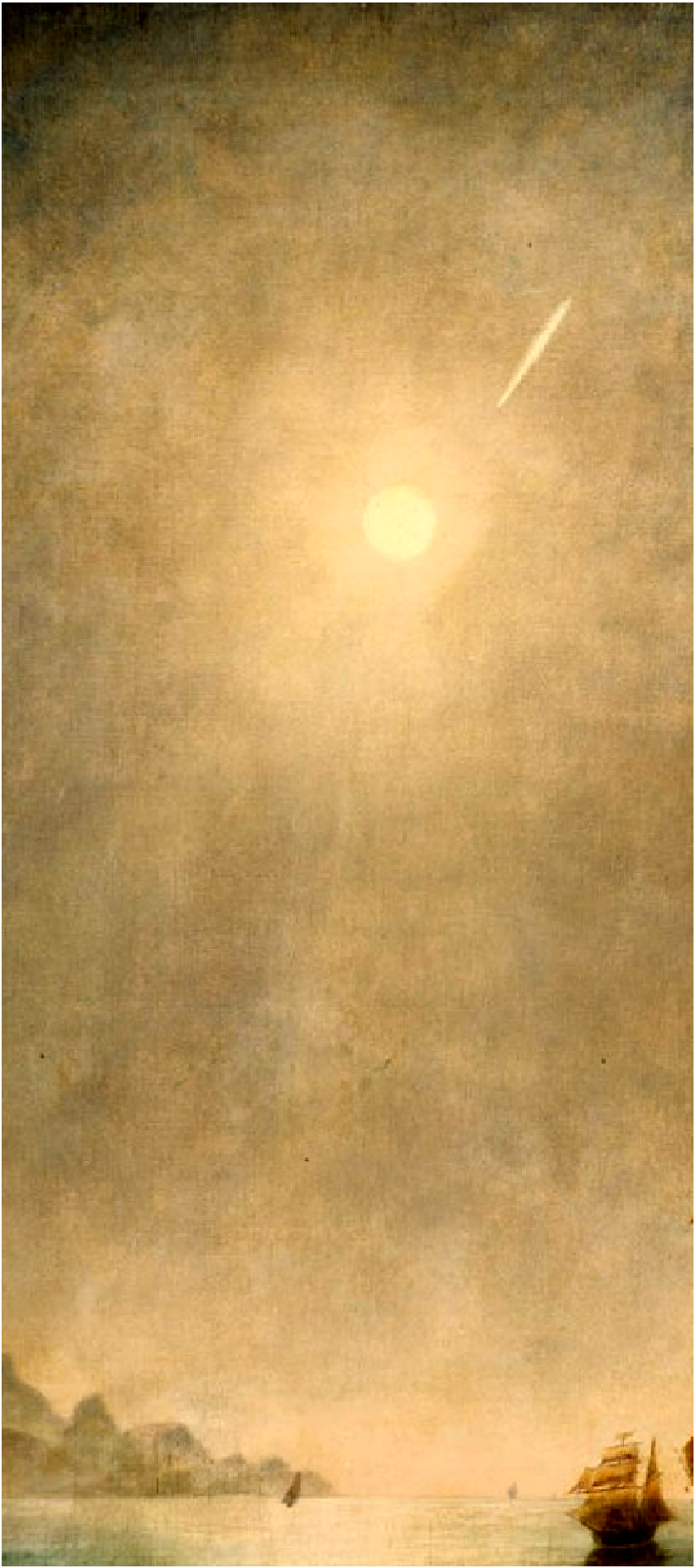}}}
\vspace{-0.05cm}
\caption{Daytime sighting of the Great March Comet of 1843 over Table Bay near
 Cape Town on 1843 Feb.\,28.20~UT, less than 30 minutes after sunrise and
 $\sim$7~hours after perihelion.  The tail is 1$^\circ$ long.  Detail of
 a painting by C.\ Piazzi Smyth, Royal~Observatory, Cape of Good Hope. (National Maritime
 Museum, London.){\vspace{0.55cm}}}
\end{figure}

Piazzi Smyth was also the author of two paintings whose details are copied
below.  The first one, made from Table Bay, just to the north of Cape Town,
is in Figure~9, showing a daytime sighting of the comet in close proximity
of the Sun.  From the distance, the comet appears to have been drawn on
February~28.20~UT, when it was 1$^\circ\!$.1 from the Sun in position angle
of 130$^\circ$. The comet then moved at a rate of 15$^\prime$ per hour.  The
straight, narrow tail, probably dominated by plasma, was 1$^\circ$ long,
pointing in position angle of 134$^\circ$.

The other of the two paintings, whose detail is reproduced in Figure~10, was
made from the area of the Signal Station, Lion's Rump, to the west of the
Royal Observatory, or from a nearby beach.  The painting shows the comet
shortly before its head set on March~4;\footnote{At the Royal Observatory
the comet set on March~4.7753~UT = 18:36.5 UT = 7:50.4 local mean time,
81~minutes after the Sun.} the comet was about 17$^\circ$ from
the Sun.  The very bright star above the comet's head was $\beta$ Ceti,
while the four stars to the right of the tail were, counterclockwise from
the bottom up, $\eta$, $\theta$, $\zeta$, and $\tau$ Ceti.  From an average
scale, derived from the distances among these stars, the tail length came
out to be 34$^\circ$ long, its position angle 106$^\circ$, essentially
equaling the direction of the prolonged radius vector.  Unquestionably,
this was a plasma tail, a conclusion implied by both its morphology, length,
and orientation.  It is noted that, incredibly, the tail was bright enough
to reflect in water.

\begin{figure} 
\vspace{0.92cm}
\hspace{-0.54cm}
\centerline{
\scalebox{1.38}{
\includegraphics{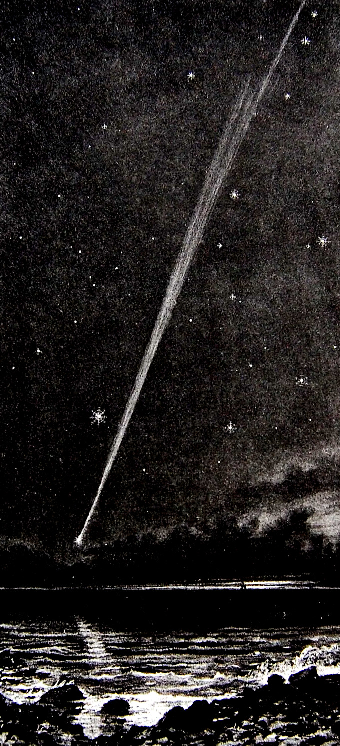}}}
\vspace{0.1cm}
\caption{The Great March Comet of 1843 in the constellation of Cetus seen from
 Cape Town in the evening of March~4, shortly before it set at 18:36~UT.
 The comet was nearly 5~days after perihelion and approximately 17$^\circ$
 from the Sun in the sky.  According to his diary, C.\ Piazzi Smyth went with
 instruments to the Signal Station on the Lion's Rump, from where  the comet
 set on the sea horizon (Warner 1980).  The painting was made either there
 or on a nearby beach.  Even though the diary does not provide information
 on the tail length on this day, from the configuration of the
 bright stars $\beta$, $\eta$, $\theta$, $\zeta$, and $\tau$ Ceti, depicted
 in the picture, I established that it amounted to $\sim$34$^\circ$, including
 the streamer at the upper end.  The tail's axis extended almost exactly in
 the antisolar direction, its position angle having equaled 106$^\circ$.
 The tail was dominated by plasma, dust may have contributed a little near
 the head.  Note that at its more distant parts the tail was double and that
 the painting shows the comet's reflection in water. (Detail of painting;
 National Maritime Museum, London.){\vspace{0cm}}} 
\end{figure}

Next, I turned to the available observations of the 1843 comet's tail
orientation and length to determine, as I did for the 1882 comet, the peak
radiation pressure acceleration $\beta_{\rm peak}$ on dust particles
located at the tail's end point and the effective time $t_{\rm ej}$ at
which they were ejected.  The results of this modeling are presented
in Table~6 for ten dates between March~8 and March~22, on which either
the coordinates of the tail's end point or the tail length and orientation
(position angle) were measured.  Nine of the ten data points were provided
by Piazzi Smyth (Warner 1980), one came from Knorre (1843).  The residuals
from two different models fitting the observations about equally well are
listed in the table, suggesting that in terms of both the peak radiation
acceleration (\mbox{$\beta_{\rm peak} \simeq 0.6\!-\!0.7$ the} solar
gravitational acceleration) and the effective ejection time
(\mbox{$t_{\rm ej} \simeq 0.2\!-\!0.3$ day} after perihelion) the 1843
sungrazer's behavior resembled that of comet Ikeya-Seki, but differed
from the 1882 sungrazer's.

In column 2 of Table 6 I present --- as in Table 1 for the 1882 comet ---
a cometocentric latitude of the Earth, which determines the foreshortening
involved in the observer's view of the distribution of dust particles in the
comet's orbital plane.  The tabulated numbers demonstrate that the geometry
of the 1843 encounter of the Earth with the Great March Comet was rather
unfavorable, as the latitude was in absolute value smaller than 10$^\circ$
on March~8 and was getting smaller with time.  At the time of observation
on March~22 the Earth was crossing the plane of the comet's orbit, so that
the observer was looking at the dust tail edge on.  At that time the
position angle of the observed tail equaled the position angle of the
projected orbital plane and provided no information whatsoever on the
distribution of dust in the plane, the residuals merely reflecting the
errors of observation.  The unfavorable geometry was in part to blame for
a very little difference in the quality of fit offered by the two tabulated
models A and B.

\begin{table*}[t] 
\vspace{0.17cm}
\hspace{-0.2cm}
\centerline{
\scalebox{1}{
\includegraphics{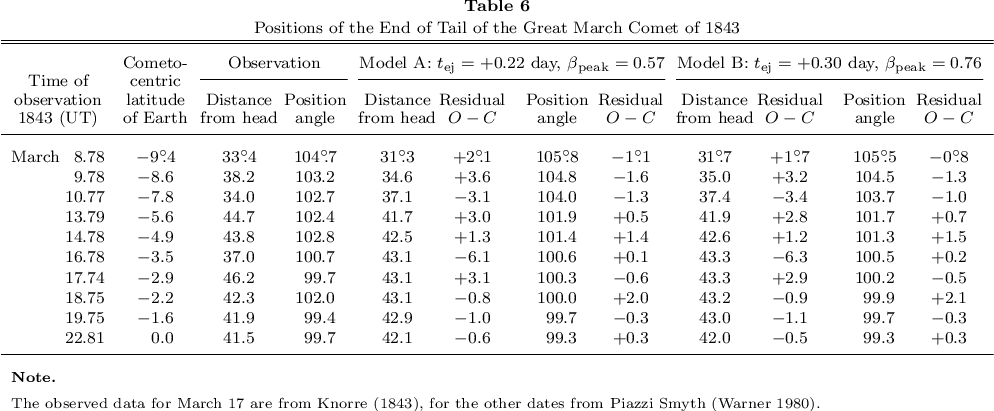}}}
\vspace{0.6cm}
\end{table*}

\subsection{Tail Length of the 1843 Sungrazer}
The tail lengths reported in accessible publications are summarized in
Table~C--1 of Appendix~C and plotted as a function of time in Figure~11.
The plotted data are compared with a model whose parameters for the dust
populating the tail's end point are about midway between those for the
models A and B in Table~6, namely, \mbox{$t_{\rm ej} = t_\pi + 0.25$ day}
and \mbox{$\beta_{\rm peak} = 0.65$ the} solar gravitational acceleration.

\begin{figure}[b] 
\vspace{0.7cm}
\hspace{-0.19cm}
\centerline{
\scalebox{0.75}{
\includegraphics{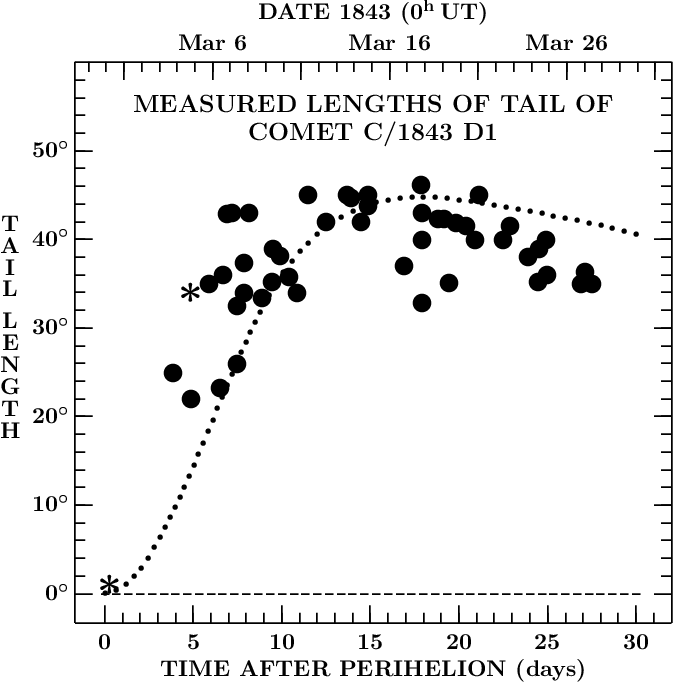}}}
\vspace{-0.05cm}
\caption{Reported lengths of the post-perihelion tail of the Great March Comet
 of 1843.  The solid circles are the entries from Table C--1, the asterisks are
 the data obtained from Figures~9 and 10, respectively.  The dotted curve shows
 a dust tail model whose end point contained particles ejected 0.25~day after
 perihelion and subjected to a radiation pressure acceleration equaling 0.65
 the solar gravitational acceleration.  Up to about 10~days after perihelion,
 the contribution from the plasma tail keeps the reported lengths above the
 curve of expected length for the dust tail.  Beyond about 20~days after
 perihelion, the reported lengths begin to drop below the curve as the
 surface brightness of the tail near its end point becomes too low to
 detect.{\vspace{-0.05cm}}}
\end{figure}

Figure 11 shows that up to about 10 days after perihelion, the substantial
contribution from the plasma tail kept the reported lengths above the curve
of expected length for the dust tail.  This was also true for the tail lengths
measured from Piazzi Smyth's paintings in Figures~9 and 10.  Beyond about
20~days after perihelion, the reported lengths began to drop below the curve,
as the surface brightness of the tail near its end point became too low to
detect.

\section{Comparison of the Three Sungrazers}
Notwithstanding, the observed tail length of the Great March Comet of 1843
exceeded the tail lengths of both the Great September Comet of 1882 and
Ikeya-Seki, as seen by inspecting Figures~5, 7, and 11.  The subject of
this section is to investigate why is that so. 

\begin{figure*} 
\vspace{0cm}
\hspace{-0.45cm}
\centerline{
\scalebox{5.44}{
\includegraphics{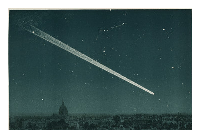}}}
\vspace{-0.05cm}
\caption{The Great March Comet of 1843 over Paris in the evening of March~19.
 The length of the tail was close to 45$^\circ$, corresponding to 130~million~km
 in space.  The Earth was about to cross the orbital plane of the comet (on
 March~22), which explains that the tail appears nearly perfectly rectilinear.
 (From Guillemin 1875).{\vspace{0.46cm}}}
\end{figure*}

Before I directly compare the three sungrazers, I should point out that the
already noted Earth's transit of the orbital plane has distinguished
the 1843 comet from the other two.  The event, which took place four weeks
after the comet had passed perihelion, at a time when the tail was still
prominent, had two implications for its appearance.  One, for several
days on either side of the transit time, the tail looked perfectly straight.
And two, because of the greater optical depth implied by the edge-on view,
the tail appeared a little longer than it otherwise would.  The well-known
sighting of the 1843 sungrazer over Paris, reproduced in Figure~12, is an
excellent example.

Because of the relative orbital positions of the Earth and
the Kreutz system's members, a post-perihelion tail displayed by a morning
sungrazer in September projects shorter\footnote{This effect has
nothing in common with foreshortening associated with the Earth's
distance from the orbit plane noted above.} than an equally long tail of
an evening sungrazer in March.\footnote{One would expect that, because of
this effect, more historical Kreutz sungrazers should have been noted in
February--March than in September--October, but the results of Hasegawa \&
Nakano's (2001) investigation do not support this inference.}  This was
the reason why the reported tail lengths of the 1843 comet exceeded
significantly those of
the 1882 comet.  The striking difference of this kind is plainly apparent
from Table~7, in which the observed tails of the three objects are
compared at 5, 15, and 25~days after perihelion and which also presents
the {\it maximum\/} reported tail lengths.  Since these were in any case
estimates, they were burdened by errors that propagated unevenly
into the true tail lengths by conversion from the observed tail lengths.
This conversion was accomplished in one of two ways.  When the contribution 
from the plasma tail was obvious, the tail was assumed to extend along the
radius vector and its length in space, $L$, was derived from the observed
length, $\ell$, with help of the well-known formula,
\begin{equation}
L = \frac{\Delta \sin \ell}{\sin (\phi \!-\! \ell)},
\end{equation}
where $\phi$ is the phase angle, Sun--comet head--Earth, and $\Delta$ is the
comet's geocentric distance.  When the observed tail was believed to be
dominated by dust, the length $L$ was derived from the equations for
dust-particle dynamics, using the parameters determined in the previous
sections.

The tabulated results suggest that the true tail lengths of the 1843 and 1882
sungrazers were comparable, but varied in a wide range, centered approximately
on 1~AU.  The tail of Ikeya-Seki was shorter than 1~AU.  The maximum lengths,
reported for Ikeya-Seki by Boethin, appear to be overestimates, especially
the higher value, which implied for the true length an unacceptably large
value, in excess of 2~AU, regardless of the model used in the conversion.
The lower value of 30$^\circ$ was closer to being in line with what one would
expect from the other numbers in the table; part of the problem was that a
relatively modest difference of 8$^\circ$ in the observed length of the tail
changed its true length by fully 1~AU.

The common trait of the tails of the three sungrazers was a major contribution
from the plasma component, determining their length in the early
post-perihelion times.  This rule was unaffected by the sodium tail,
detected in comet Ikeya-Seki.  As the objects were receding from the
Sun, the plasma share was gradually disappearing, until the tails
eventually consisted of pure dust.

The sungrazers also displayed similar tail-length variations with time.
Shortly after perihelion  the tail was growing longer, reached a maximum,
and then its length subsided with increasing geocentric distance.
Comparison with dust-tail models suggested that in the early
post-perihelion period of time the observed tail length exceeded the
model prediction, obviously because of the presence of the plasma
tail, while months after perihelion the predicted length exceeded
the observed length, apparently an effect of the decreasing surface
brightness.

\begin{table}[t] 
\vspace{0.14cm}
\hspace{-0.18cm}
\centerline{
\scalebox{0.99}{
\includegraphics{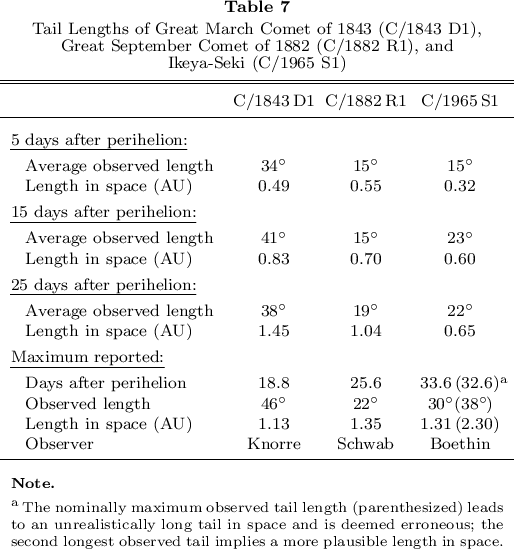}}}
\vspace{0.7cm}
\end{table}

The dust tails of the Great March Comet and Ikeya-Seki were also alike
in that their end points were populated by the ejecta that left the
nucleus a small fraction of a day after perihelion and were subjected
to radiation pressure accelerations not exceeding 0.6--0.7 the solar
gravitational acceleration.  On the other hand, the dust tail of the
Great September Comet contained dust particles that appeared to move
under radiation pressure accelerations of up to about twice the solar
gravitational acceleration and were ejected from the nucleus days
after perihelion, possibly an effect of the extensive fragmentation
of the nucleus.  This sungrazer also displayed a peculiar feature
of a comet in a comet, believed to be a result of near-perihelion
disintegration of a companion that had traveled in the same orbit
about the Sun for centuries, reaching perihelion nearly 7~days after
the main nucleus.

\section{Conclusions}
This study was prompted by the important role of the post-perihelion
tails in the first sightings of historical appearances of the
potential Kreutz sungrazers, implied by the results of Hasegawa \&
Nakano's (2001) investigation.  The objectives are to find out whether
it is the plasma or dust component that determines a sungrazer's
apparent tail length; variations in the tail length with time
and as a function of season; the dynamical properties of the
particulate material in the dust tails; peculiar tail features;
and the correlation between the observed tail length and true
tail length in interplanetary space.

As the pre-perihelion tail completely sublimates at perihelion, the
post-perihelion tail development begins from scratch.  Under these
conditions it is to be expected that shortly after perihelion the tail
gradually increases in length, starting from nil.  Yet, the rate of this
increase often appears to be phenomenally rapid, as illustrated by the
Great March Comet of 1843:\ Less than five days after perihelion its
tail length on a painting by Piazzi Smyth is estimated at 34$^\circ$
(Figure~10).  Gigantic dimensions of the plasma tail are made possible
by enormous accelerations on CO$^+$ and other ionized molecules.
The dust tail is at these times much shorter, depending on the peak
radiation pressure acceleration that sometimes does not exceed 0.6--0.7
the solar gravitational acceleration.  The accelerations in the plasma
tail are two or more orders of magnitude higher.  
 
The radiation pressure acceleration is one of two fundamental parameters
that govern the motion of a dust particle through the tail.  The other is
the time of ejection, while the ejection velocity is also a contributing
factor.  When the subject of interest is the tail length, the objective
is to find the critical parameters of the ejecta at the tail's end point.
The tail end of a Kreutz sungrazer weeks after perihelion is usually
populated by particles ejected a fraction of the day after perihelion.
This is what I find for both Ikeya-Seki and the 1843 sungrazer; the 1882
sungrazer turns out to be a more complicated case.  The big help in my
effort to solve the problem has been Gill's (1882) photograph and
Schmidt's (1882) study of the $\alpha\,^\prime$ feature seen both
visually and on the plate.  Even though Schmidt's treatment was tentative,
my reanalysis of the motion of the feature --- located near the end
point of the tail, but far from its sharp edge (populated by the dust
subjected to the highest accelerations) --- shows that it consisted
of grains released later after perihelion than the ejecta at the
tail's end point of the two other comets.

Although the data on the tail's end point for the 1882 sungrazer are poor,
available information does seem to suggest that the dust tail of this
object differed from those of Ikeya-Seki and the 1843 sungrazer, possibly
because of the extensive fragmentation of its nucleus near perihelion.
The continuing high activity of the nuclear fragments could have made
the difference.
 
The 1882 comet was unique in displaying the bizarre feature of a comet
in a comet, brought apparently about by the sudden disintegration of
a distant companion nucleus that long followed the main comet in the
orbit, passing perihelion nearly 7~days later.

I am unable to confirm a statement often found in popular publications
on comets that the true tail length of the Great March Comet of 1843 was
2~AU long, longer than any other comet.  Nonetheless, the derived length
of nearly 1.5~AU 25~days after perihelion is also respectable.  The Earth
crossing the comet's orbit plane may have contributed to the reported
tail lengths at the time.

An interesting result is a major difference in the degree of foreshortening
affecting the post-perihelion tails of the morning sungrazers in
September--October and the evening sungrazers in February--March.  All
other things equal, the evening objects have the projected tails
substantially longer than the morning objects.

Addressing finally the initial issue that provoked~this investigation,
there is no doubt that the discoveries~of the historical appearances
of potential Kreutz sungrazers, made mostly in the first two weeks
after perihelion, as follows from Hasegawa \& Nakano's (2001) list,
were achieved because of these objects' prominent tails.  I also
conclude that the tail length was not a critical parameter that
facilitated detection; if it were, more potential sungrazers would have
been sighted in February--March than in September--October.  They were
not. \\[-0.3cm]

\begin{center}
{\large \bf Appendix A} \\[0.3cm]
SUMMARY OF\\POST-PERIHELION TAIL OBSERVATIONS OF\\COMET IKEYA-SEKI
 (C/1965 S1)
\end{center}

\vspace{-0.2cm}
This appendix provides lists of the observations of the apparent tail lengths
in Table~A--1 and of the position angles in Table A--2.\\[-0.05cm]

\begin{table*}[t] 
\vspace{0.17cm} 
\hspace{0cm}
\centerline{
\scalebox{1}{
\includegraphics{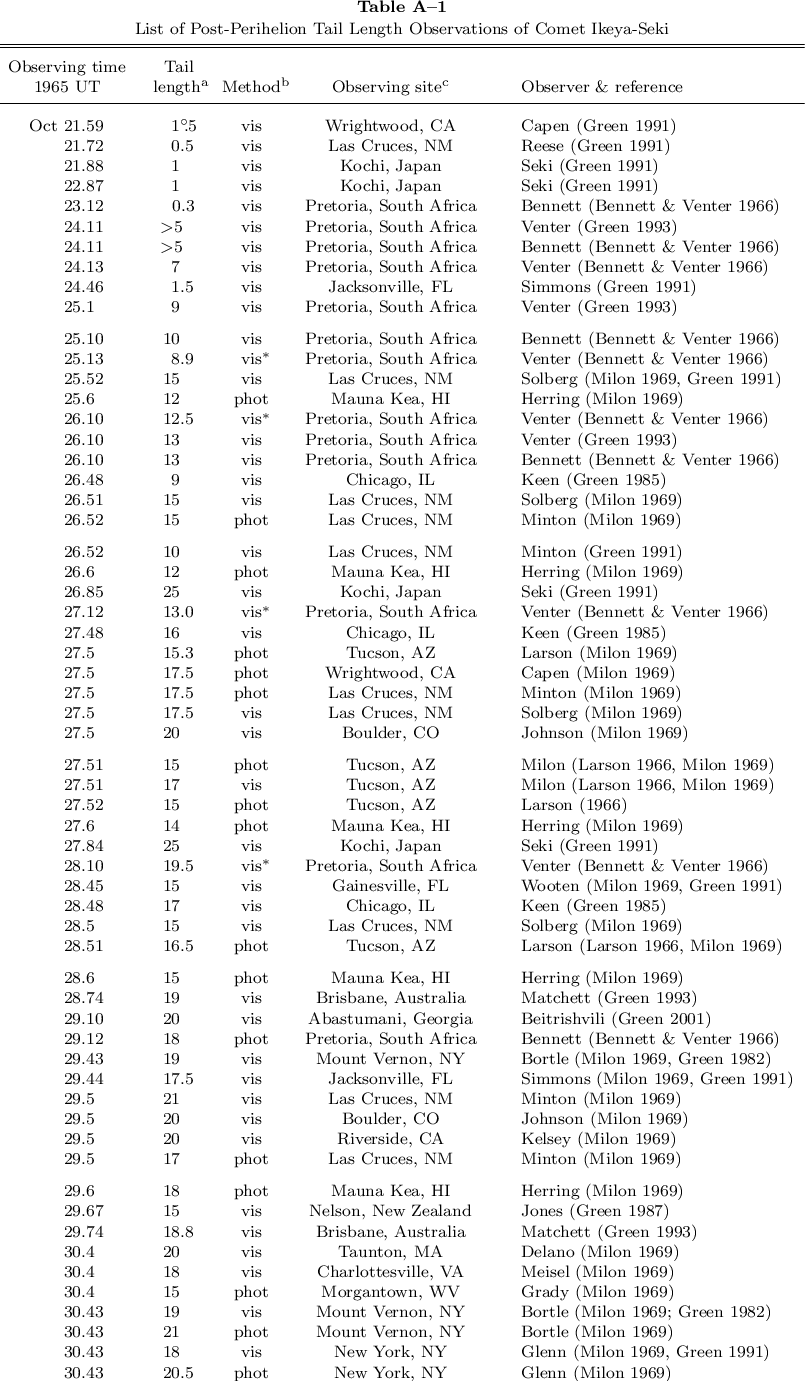}}}
\end{table*}
\begin{table*}[t] 
\vspace{0.17cm} 
\hspace{0cm}
\centerline{
\scalebox{1}{
\includegraphics{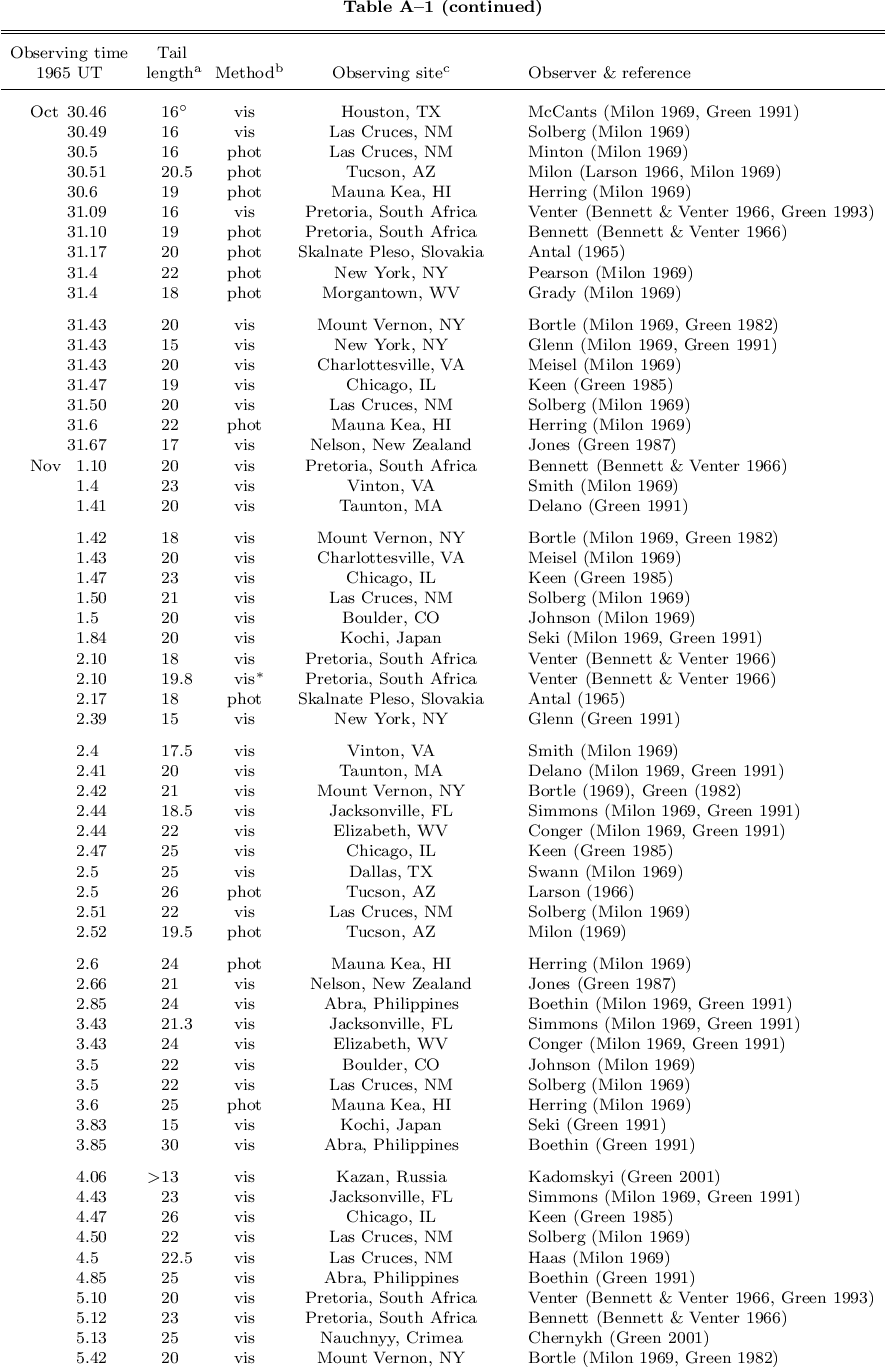}}}
\end{table*}

\vspace{-0.3cm}

\begin{table*}[t] 
\vspace{0.17cm} 
\hspace{0cm}
\centerline{
\scalebox{1}{
\includegraphics{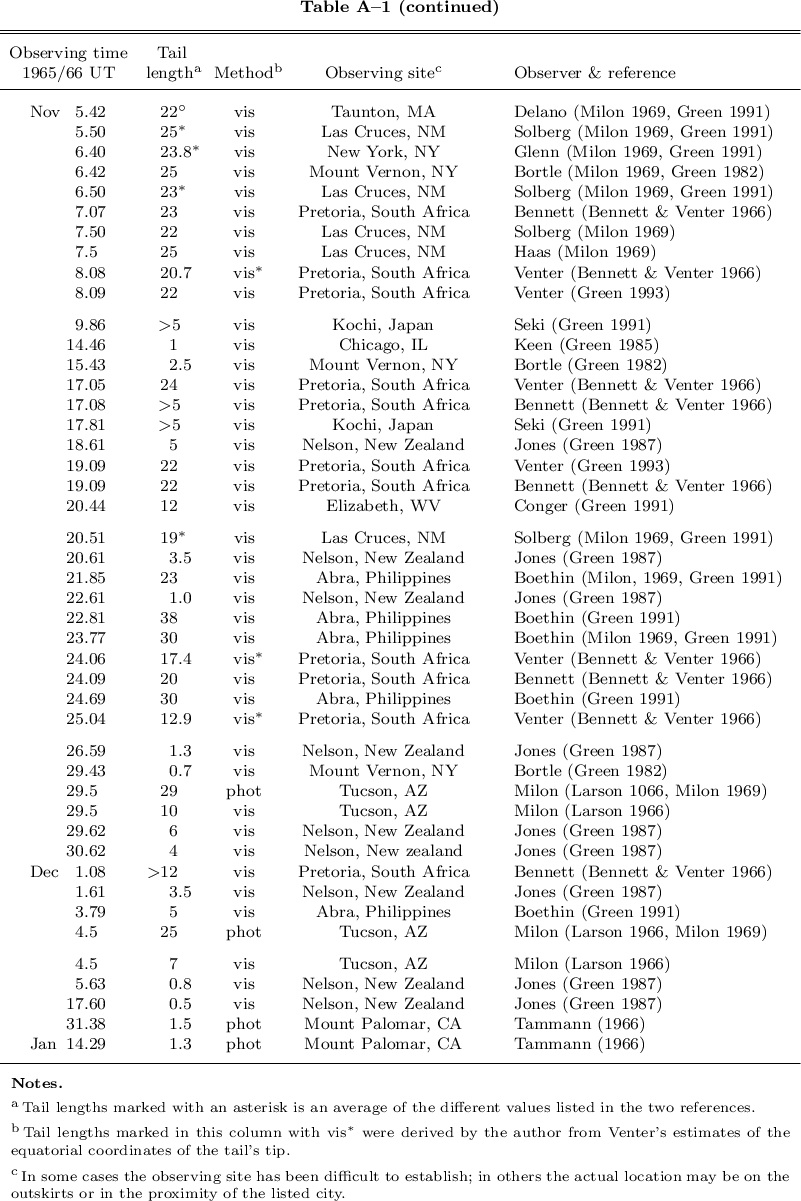}}}
\end{table*}
\begin{table*}[t] 
\vspace{0.17cm}
\hspace{0cm}
\centerline{
\scalebox{1}{
\includegraphics{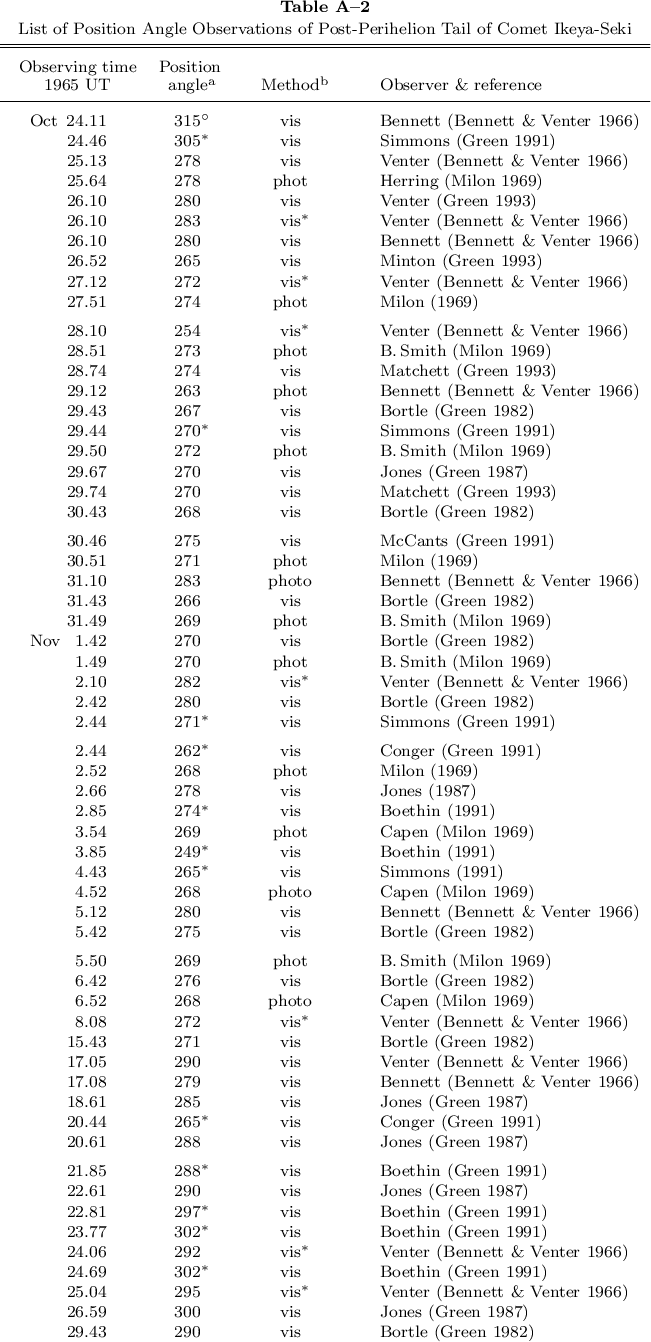}}}
\end{table*}

\begin{table*}[t] 
\vspace{0.17cm}
\hspace{0cm}
\centerline{
\scalebox{1}{
\includegraphics{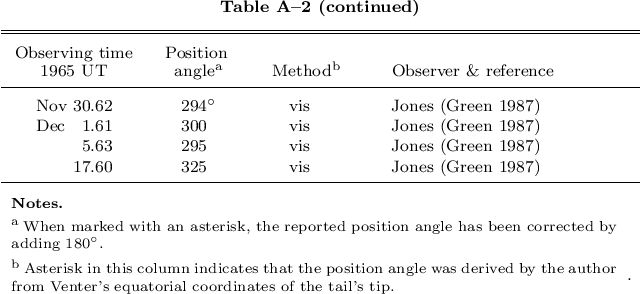}}}
\vspace{-4cm}
\end{table*}

\begin{table*}
\begin{center}
{\large \bf Appendix B} \\[0.2cm]
{\normalsize SUMMARY OF\\POST-PERIHELION TAIL OBSERVATIONS OF\\[0.08cm]
 GREAT SEPTEMBER COMET OF 1882  (C/1882 R1)}
\end{center}
{\hspace{4.65cm}}{\parbox{8.7cm}{\normalsize This appendix provides a list
of the apparent tail length
{\hspace{4.65cm}}observations in Table~B--1.}}
\begin{center}
{\Huge \boldmath $\Downarrow$}
\end{center}
\end{table*}

\begin{table*}[t] 
\vspace{-2.4cm}
\hspace{1cm}
\centerline{
\scalebox{1}{
\includegraphics{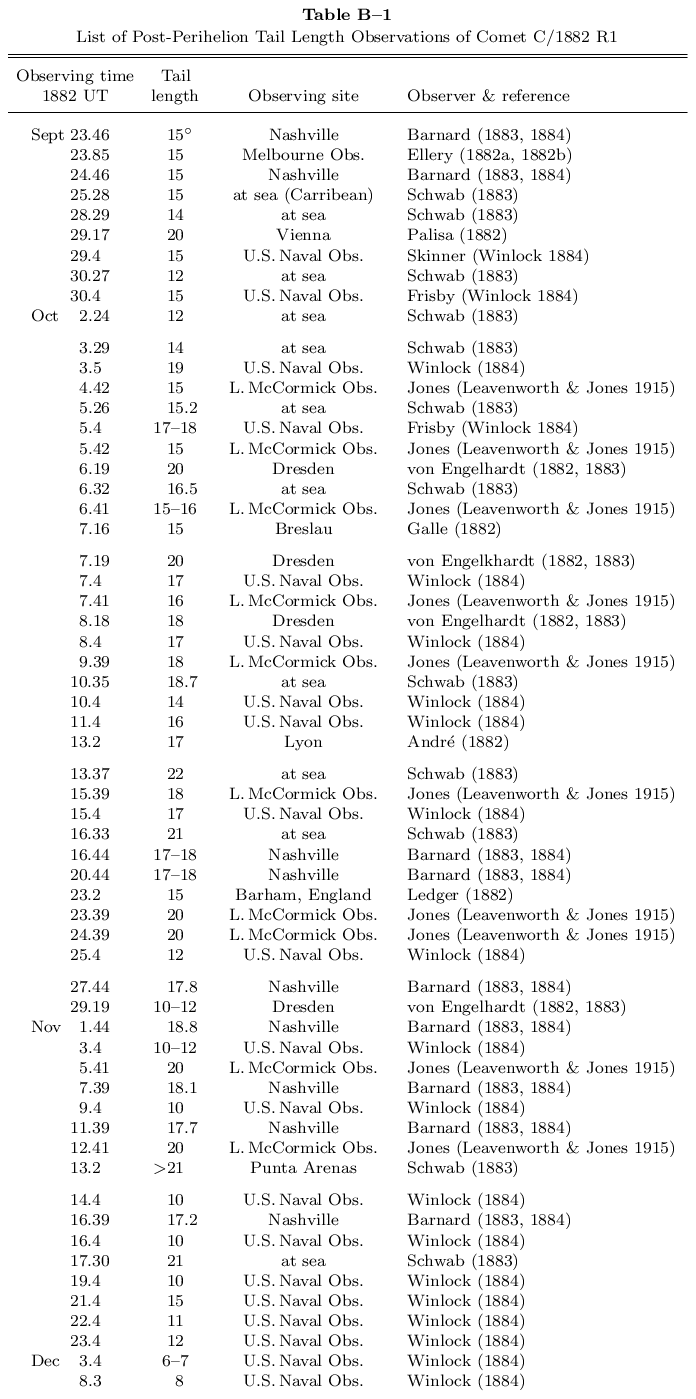}}}
\end{table*}

\begin{table*}[t] 
\vspace{-2.4cm}
\hspace{0.65cm}
\centerline{
\scalebox{1}{
\includegraphics{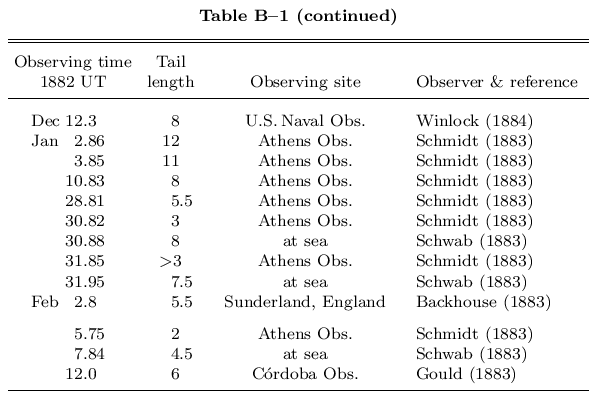}}}
\vspace{-25cm}
\end{table*}
\begin{table*}
\begin{center}
{\large \bf Appendix C} \\[0.2cm]
{\normalsize SUMMARY OF\\POST-PERIHELION TAIL OBSERVATIONS OF\\[0.08cm]
 GREAT MARCH COMET OF 1843 (C/1843 D1)}
\end{center}
{\hspace{4.65cm}}{\parbox{8.7cm}{\normalsize This appendix provides a
list of the apparent tail length
{\hspace{4.65cm}}observations in Table~C--1.}}
\begin{center}
{\Huge \boldmath $\Downarrow$}
\end{center}
\end{table*}

\begin{table*}[t] 
\vspace{-2.4cm}
\hspace{1cm}
\centerline{
\scalebox{1}{
\includegraphics{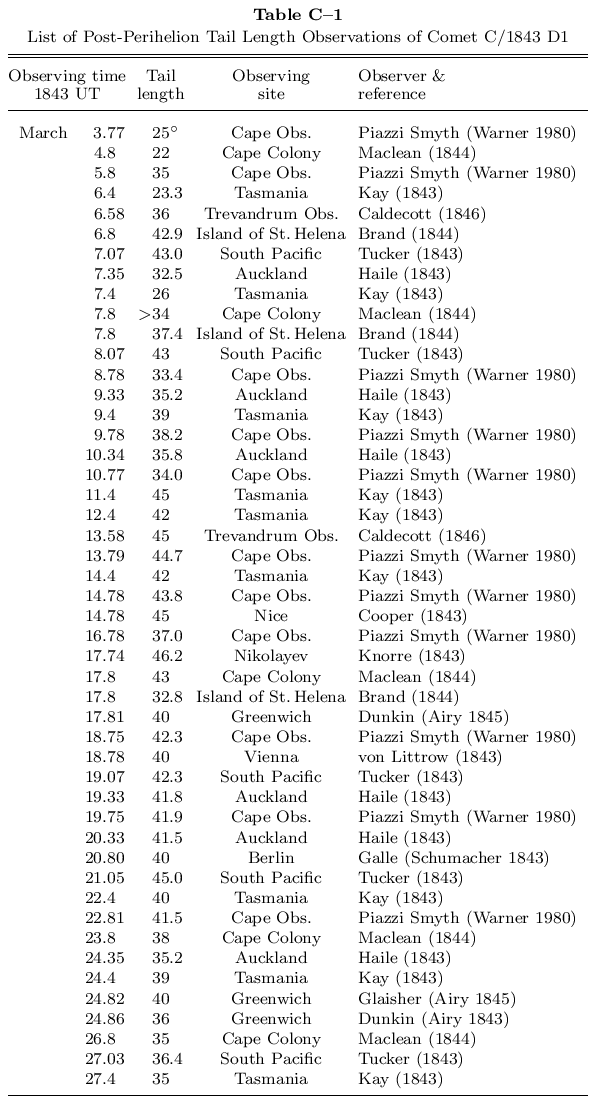}}}
\end{table*}
%

%
\begin{table*}[t]
\vspace{0.3cm}
\hspace{-0.3cm}
\leftline{
\scalebox{1.01}{
\includegraphics{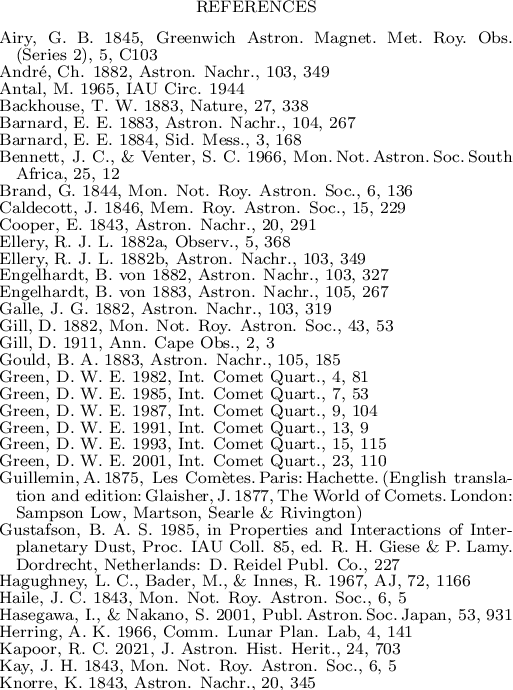}}}

\vspace{-11.82cm}
\hspace{0.2cm}
\rightline{
\scalebox{1.006}{
\includegraphics{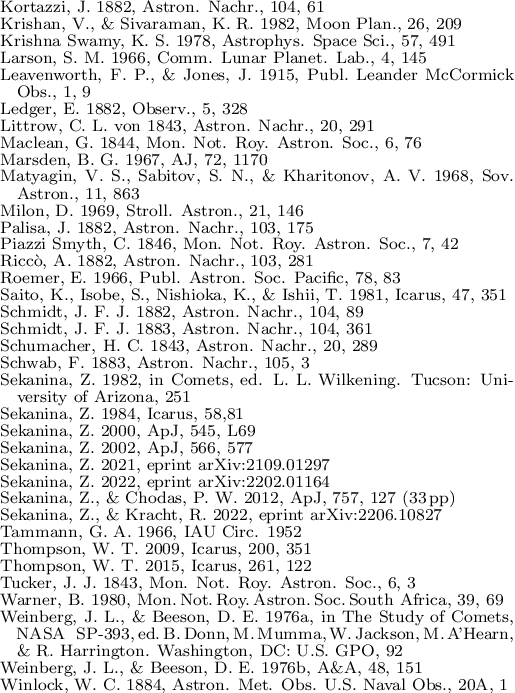}}}
\end{table*}
\end{document}